\documentclass[%
 reprint,
 superscriptaddress,
 amsmath,amssymb,
 aps,
prapplied,
doi=true,
floatfix,
]{revtex4-1}

\usepackage[colorlinks=true,linkcolor=blue,urlcolor=blue,citecolor=blue]{hyperref}

\usepackage{graphicx}
\usepackage{bm,upgreek}

\usepackage{overpic}
\usepackage{comment}
\usepackage{xfrac}
\usepackage{xcolor}
\usepackage{dblfloatfix}
\usepackage{mathtools}
\usepackage[mathscr]{euscript}

\usepackage[export]{adjustbox}
\usepackage{amssymb}
\usepackage{relsize}

\let\savecorresponds\corresponds
\let\corresponds\relax
\usepackage{mathabx}
\let\corresponds\savecorresponds

\definecolor{dandelion}{rgb}{0.94, 0.88, 0.19}
\definecolor{thepurple}{rgb}{0.65, 0.24, 0.59}
\definecolor{theorange}{rgb}{0.95, 0.40, 0.13}
\definecolor{thegreen}{rgb}{0.05, 0.50, 0.25}
\definecolor{mathematicagreen}{rgb}{0.0, 0.80, 0.0}
\definecolor{igoraquamarine}{rgb}{0.0, 0.6, 0.6}
\definecolor{igorgreen}{rgb}{0.0, 0.6, 0.0}
\definecolor{springgreen}{rgb}{0.0, 0.8, 0.0}
\definecolor{skyblue}{rgb}{0.0, 0.6, 1.0}
\definecolor{black}{rgb}{0.0 0.0 0.0}

\begin{document}

\preprint{APS/123-QED}

\title{Photoluminescence decomposition analysis: a technique to characterize NV creation in diamond}

\author{Scott T. Alsid}
\affiliation{Lincoln Laboratory, Massachusetts Institute of Technology, Lexington, Massachusetts 02421, USA}
\affiliation{Department of Nuclear Science and Engineering, Massachusetts Institute of Technology, Cambridge, Massachusetts 02139, USA}
\author{John F. Barry}
\email[Corresponding author.\\]{john.barry@ll.mit.edu}
\affiliation{Lincoln Laboratory, Massachusetts Institute of Technology, Lexington, Massachusetts 02421, USA}
\author{Linh M. Pham}
\affiliation{Lincoln Laboratory, Massachusetts Institute of Technology, Lexington, Massachusetts 02421, USA}
\author{Jennifer M. Schloss}
\affiliation{Department of Physics, Massachusetts Institute of Technology, Cambridge, Massachusetts 02139, USA}
\affiliation{Center for Brain Science, Harvard University, Cambridge, Massachusetts 02138, USA}
\author{Michael F. O'Keeffe}
\affiliation{Lincoln Laboratory, Massachusetts Institute of Technology, Lexington, Massachusetts 02421, USA}
\author{Paola Cappellaro}
\affiliation{Department of Nuclear Science and Engineering, Massachusetts Institute of Technology, Cambridge, Massachusetts 02139, USA}
\author{Danielle A. Braje}
\affiliation{Lincoln Laboratory, Massachusetts Institute of Technology, Lexington, Massachusetts 02421, USA}

\date{\today}

\begin{abstract}
Treatment of lab-grown diamond by electron irradiation and annealing has enabled quantum sensors based on negatively-charged nitrogen-vacancy (NV$^\text{-}$) centers to demonstrate record sensitivities. 
Here we investigate the irradiation and annealing process applied to 28 diamond samples using a new ambient-temperature, all-optical approach. As the presence of the neutrally-charged nitrogen-vacancy (NV$^\text{0}$) center is deleterious to sensor performance, this photoluminescence decomposition analysis (PDA) is first employed to determine the concentration ratio of NV$^\text{-}$ to NV$^0$ in diamond samples from the measured photoluminescence spectrum. The analysis hinges on (i) isolating each NV charge state's emission spectrum and (ii) measuring the NV$^\text{-}$ to NV$^0$ emission ratio, which is found to be 2.5$\pm$0.5 under low-intensity 532 nm illumination. Using the PDA method, we measure the effects of irradiation and annealing on conversion of substitutional nitrogen to NV centers. Combining these measurements with a phenomenological model for diamond irradiation and annealing, we extract an estimated monovacancy creation rate of $0.52\pm 0.26$ cm$^{\text{-1}}$ for 1 MeV electron irradiation and an estimated monovacancy diffusion coefficient of 1.8 nm$^2$/s at 850~$^\circ$C. Finally we find that irradiation doses $\gtrsim 10^{18}$ e$^\text{-}$/cm$^2$ deteriorate the NV$^\text{-}$ decoherence time $T_2$ whereas $T_1$ is unaffected up to the the maximum investigated dose of $5\times 10^{18}$ e$^\text{-}$/cm$^2$.
\end{abstract}

\maketitle

\section{\label{sec:intro}Introduction}

The negatively-charged nitrogen vacancy (NV$^\text{-}$) center, with its rich spin dynamics even at room temperature, has captured the interest of the sensor community. Importantly, quantum sensors based on NV$^\text{-}$ ensembles in diamond require efficient conversion of substitutional nitrogen to NV$^\text{-}$ centers to achieve peak performance~\cite{acosta2009diamonds, Wolf2015, Kleinsasser2016}. One common approach to creating NV$^\text{-}$-rich diamond material involves electron-irradiating and subsequently annealing nitrogen-doped diamond~\cite{Clevenson2015,Wolf2015,Barry2016,Chatzidrosos2017}. However, maximizing the NV$^\text{-}$ population while minimizing unwanted defects (e.g., neutral NV$^0$ centers, vacancy complexes, and other paramagnetic impurities) is critically dependent on employing proper irradiation and annealing processes.

Although high-quality diamond processing studies are available~\cite{Hartlandthesis2014} to guide post-growth treatment, literature discrepancies (see Appendix Sec.~\ref{Appendix:IrradiationStudy} and Refs.~\cite{Collins2009,davies1999current}) and a shortage of systematic studies impede full understanding of these post-growth treatments affect defect composition. In particular, variation in the base diamond material employed in the literature studies constitutes a prominent and continuing obstacle to optimizing treatment parameters. For example, diamond origin (natural or synthetic), growth method (CVD or HPHT), nitrogen concentration~\cite{davies1992vacancy,collins2003production,iakoubovskii2005evidence}, lattice defect density, impurity concentration~\cite{collins2007optical}, and strain~\cite{collins2007optical} have all been shown to affect the irradiation and annealing process. As diamond synthesis capabilities develop, producing material with a range of compositions, a comprehensive study of irradiation and annealing will enable targeted diamond engineering for diverse quantum sensing applications~\cite{Choi2017,Wolf2015,Zheng2018zerofieldmagnetometry}.

To fine-tune the diamond treatment process, accurate measurements of the effects of irradiation and annealing on key diamond material properties are needed. In particular, optimizing NV-diamond quantum sensors necessitates post-treatment characterization of (i) the conversion efficiency from nitrogen to NV centers, (ii) the NV$^\text{-}$ to NV$^0$ charge state ratio, and (iii) the relevant NV$^\text{-}$ spin relaxation times. Such measurements can also provide physical insight into the vacancy creation and diffusion dynamics central to formation of NV$^\text{-}$ centers. 
Here we present an all-optical, ambient-temperature technique for rapidly characterizing the NV$^\text{-}$ to NV$^0$ charge state ratio and total NV concentration. This technique provides an appealing alternative to slower conventional methods, such as electron paramagnetic resonance (EPR) and ultraviolet-visible (UV-Vis) spectroscopy. Further, the technique is suitable for analyzing NV-doped layers or other small ensembles, regimes typically precluded from EPR or UV-Vis analysis by signal-to-noise requirements. Additionally, accurate charge state ratio measurements hinge on knowledge of the photoluminescence (PL) rate of NV$^\text{-}$ relative to NV$^0$ for equal concentrations. Here we determine for the first time this PL ratio under low-intensity 532 nm excitation.

We demonstrate the presented technique's utility in a systematic study of irradiation and annealing on 28 diamonds. We focus on a diamond base material that is commercially available and relevant to many quantum sensing applications~\cite{Wolf2015,Zheng2018zerofieldmagnetometry}. To support additional analysis of our measurements, we introduce a phenomenological model for the irradiation and annealing process and extract an estimated irradiation-induced vacancy creation rate of $0.52\pm 0.26$ cm$^{-1}$ for 1 MeV electron irradiation. From the model, we also determine an estimated monovacancy diffusion coefficient of 1.8 nm$^2$/s at 850~$^\circ$C. Further, we examine the effects of irradiation and annealing on the NV$^\text{-}$ coherence time $T_2$ and longitudinal relaxation time $T_1$ in the employed base material. 
Finally, we comment on applications of the methods and analysis detailed in this work to furthering NV physics research and diamond materials science. Example near-term studies include characterization of NV$^0$ optical dynamics and investigation into the effects of base material variation on vacancy migration.

\section{Diamond Material Optimization}

The presence of various electronic spin species in diamond (e.g., NV$^\text{-}$, NV$^0$, substitutional nitrogen N$_\text{S}$, and other paramagnetic impurities) can limit the achievable sensitivity of an NV-diamond quantum sensor; this effect motivates the careful engineering of diamond material to contain optimal concentrations of each of these spin defects. In particular, spin defect concentrations affect three key parameters that can limit sensitivity: measurement time, measurement contrast, and number of photons collected per measurement. For example, measurement time is limited by the dephasing, coherence, or relaxation times of the NV$^{\text{-}}$ spins, which in turn are typically limited by other spin species generating magnetic noise in the vicinity of the sensing NV$^{\text{-}}$ centers. Measurement contrast can be degraded by the presence of non-information-containing background fluorescence emitters like NV$^{0}$.
The number of photons collected per measurement is dictated by the optical collection efficiency of the experimental apparatus and ultimately by the number of NV centers probed during the measurement. 

In essence, the sensitivity of an NV-diamond quantum sensor can generally be enhanced by maximizing the number of NV$^\text{-}$ centers while minimizing the presence of all other spin defects. Since nitrogen is an integral constituent of the NV$^\text{-}$ center, nitrogen-containing defects often dominate in diamond material employed for NV-ensemble-based sensors. Thus, optimal values of the NV conversion efficiency $\chi \equiv [\text{NV}^\text{T}]/[\text{N}^\text{T}]$ are desired to achieve optimal sensitivity in these devices. Here $[\text{NV}^\text{T}]$ indicates the total NV population, summed over all charge states, and $[\text{N}^\text{T}]$ denotes the total nitrogen concentration in all forms including all charge states (e.g., $\text{N}_\text{S}^0,\;\text{N}_\text{S}^+,\;\text{NV}^\text{0},\;\text{NV}^\text{-}$, etc.).

Note that there is an important distinction between the NV$^{\text{-}}$ centers, which are the fundamental sensing qubits that comprise the NV-diamond quantum sensor, and NV$^0$ centers. The presence of NV$^{\text{0}}$ degrades sensor performance via several mechanisms. First, the photoluminescence from NV$^0$ spectrally overlaps with that of NV$^\text{-}$, contributing background signal and thereby reducing measurement contrast. This effect is particularly deleterious when [NV$^\text{0}$] $\gtrsim$ [NV$^\text{-}$]~\cite{Manson2005}. Additionally, the NV$^0$ center's electronic $S=\frac{1}{2}$ paramagnetic ground state~\cite{Felton2008} may increase magnetic noise in the diamond and thereby degrade NV$^{\text{-}}$ spin-coherence times. 
For these reasons, maximization of the charge state efficiency $\zeta \equiv [\text{NV}^{\text{-}}]/[\text{NV}^\text{T}]$ is also desirable. 

A simple model suggests NV$^\text{-}$ formation requires two nitrogen atoms and a vacancy. One nitrogen atom captures the vacancy to form the NV center, while the other nitrogen donates a valence electron to negatively charge the NV center~\cite{Felton2009}. For typical CVD diamonds, such as those employed in this work, the native grown-in vacancy concentration is too low to achieve optimal N$^\text{T}$-to-NV$^\text{T}$ conversion efficiency $\chi \sim 1/2$. One straightforward approach to improve $\chi$ in these diamonds is to perform electron irradiation to create vacancies in the crystal lattice. A subsequent annealing step mobilizes the vacancies and enables their capture by N$_\text{S}$ defects to form NV centers. Achieving the optimal spin defect composition in the diamond following this two-step process is critically dependent on optimizing irradiation and annealing parameters. 
For example, too low an irradiation dose will introduce too few vacancies relative to N$_{\text{S}}$ defects, limiting the attainable post-anneal NV conversion efficiency $\chi$, which in turn limits the number of NV$^\text{-}$ sensors available to perform measurements. Furthermore, the remaining under-utilized paramagnetic N$_\text{S}$ defects can degrade NV$^\text{-}$ spin-coherence times. On the other hand, too high an irradiation dose will introduce too many vacancies relative to N$_{\text{S}}$, resulting in overproduction of undesirable neutral NV$^0$ centers~\cite{Mita1996} and thereby degrading the charge state efficiency $\zeta$.

Proper selection of irradiation energy and dose, annealing time and temperature, and other processing parameters is imperative for achieving NV-diamond optimized for high-sensitivity quantum sensing. 
Consequently, quantitative characterization of key material properties, such as the NV$^\text{-}$ to NV$^0$ charge state ratio, is likewise important in determining these optimal processing parameters.
Conventional techniques for quantitatively characterizing relevant spin defect concentrations include ultraviolet-visible (UV-Vis) spectroscopy, which measures [NV$^{0}$] and [NV$^{\text{-}}$],
and electron paramagnetic resonance (EPR) spectroscopy, which measures [NV$^{\text{-}}$] and [N$_\text{S}$]. Unfortunately, UV-Vis spectroscopy is performed at cryogenic temperatures and thus requires additional time for thermal cycling. EPR spectroscopy of typically small diamond volume likewise requires significant averaging time. Since both of these conventional techniques typically take $\gtrsim$1 hour to perform measurements on a single sample, characterizing the large number ($\gtrsim$10) of samples typically involved in optimizing processing parameters can be quite onerous.
Furthermore, both techniques employ specialized equipment, which is not readily accessible in many NV-diamond research labs.

\section{Experimental design}
This study employs 28 commercially available diamonds (Element Six, 145-500-0274-01) grown via chemical vapor deposition (CVD). The chip-sized samples are 2.6 mm $\times$ 2.6 mm $\times$ 0.3 mm, with ${<}100{>}$ edges and $\{100\}$ face orientations. The diamond material contains carbon in natural isotopic abundance (1.1\% $^{13}$C, 98.9\% $^{12}$C) and exhibits a mean total nitrogen concentration [N$^\text{T}$] $\sim$ 118 ppb (see Appendix Sec.~\ref{Appendix:DiamondConcentration}). Prior to irradiation, the samples were pre-annealed at 850~$^\circ$C for one hour in a 760 Torr nitrogen atmosphere.

Diamond irradiation is performed at a commercial facility (Prism Gem LLC) using a 1 MeV electron beam with a 1 cm nominal diameter. Diamonds are placed on a water-cooled plate maintained at a temperature 66~$^{\circ}$C or below; the diamonds are estimated to remain below 120~$^\circ$C during irradiation, suggesting that interstitial-vacancy recombination during irradiation can be neglected~\cite{newton2002recombination}. Twenty-five diamonds received doses systematically varied between $1 \times 10^{15}$ e$^\text{-}$/cm$^2$ and $5 \times 10^{19}$ e$^\text{-}$/cm$^2$; the other three diamonds remained un-irradiated and were retained as controls. Nineteen of the irradiated diamonds were subsequently annealed at 850~$^{\circ}$C in a 760 Torr nitrogen atmosphere for one hour at the same commercial facility. The other six irradiated diamonds and a subset of the 850~$^{\circ}$C annealed diamonds were then annealed at 1250~$^{\circ}$C under 100 Torr hydrogen gas for one hour. Each diamond's specific irradiation and annealing parameters are given in the Appendix Sec.~\ref{Appendix:DiamondIrradiation}.\par

Diamond characterization is performed on \textcolor{black}{NV-ensembles using} a custom-built confocal microscope with attached spectrometer (See Appendix Sec.~\ref{Appendix:ConfocalMicroscope}). While this confocal microscope follows standard designs~\cite{Childress2006} in most ways, a notable modification allows enhanced functionality for NV ensemble excitation power/intensity studies. Specifically, the microscope employs a detection volume centered within a much larger excitation volume, ensuring all NV centers within the detection volume receive similar optical illumination intensity (see Appendix Sec.~\ref{Appendix:ExcitationCollectionVolume}, Fig.~\ref{excitationcollectionprofiles}). Without this modification, the intensity gradient associated with the wings of a Gaussian excitation profile can complicate interpretation of intensity-induced dynamics, such as photoionization and charge conversion.

A spectrometer attached to the confocal microscope allows spectral characterization of the photoluminescence (PL) light from 560 nm to 818 nm. This range allows collection of nearly all emitted PL light from NV$^\text{-}$ and NV$^0$; analysis of PL data in the literature from other groups~\cite{Aslam2013,Kato2013,Schreyvogel2014,Zhu2011,Fraczek2017,Monticone2013,Kehayias2013,Doi2014,Grotz2012,manson2018nv} suggests this wavelength range covers $\gtrsim\!99\%$ of NV$^\text{-}$ and NV$^0$ photoluminescence.

\section{Charge state photoluminescence study}
\label{ChargeStateSection}

In this section, we present our method for quantitatively determining the charge state efficiency $\zeta$ using PL spectroscopy. 
Note that, historically, approximations of $\zeta$ have been extracted from PL spectra~\cite{acosta2009diamonds} using the Debye-Waller factor $\tilde{S}$~\cite{Alkauskas2014,Davies1974} (see Appendix Sec.~\ref{Appendix:HuangRhysFactor}).
However, this method is non-ideal because it results in large uncertainties in the extracted $\zeta$ and does not account for the difference in PL emission of the different charge states. We employ an alternative method for isolating the PL contributions due to NV$^\text{-}$ and NV$^0$ for a given PL spectrum. We then use this alternative decomposition method to determine the ratio of PL light produced by a single NV$^\text{-}$ to that of a single NV$^0$ per unit time. The charge-state photoluminescence ratio of $\text{NV}^\text{-}$ relative to $\text{NV}^\text{0}$ is denoted $\kappa_\lambda$ for optical excitation at wavelength $\lambda$ (in this work, $\lambda=532$ nm) Using laser intensities where NV saturation effects can be neglected, the concentration of each charge state is then proportional to its respective PL when correcting for $\kappa_\text{532}$.  Therefore the charge state efficiency may be re-written as
\begin{equation}\label{eqn:chargestateefficiency}
\zeta \equiv\frac{[\text{NV}^{\text{-}}]}{[\text{NV}^\text{T}]}= \frac{[\text{NV}^{\text{-}}]}{[\text{NV}^{\text{-}}]+[\text{NV}^0]} = \frac{c_{\text{-}}}{c_{\text{-}}+\kappa_\text{532}c_0},
\end{equation}
where $c_{\text{-}}$ and $c_0$ denote the fractional contributions to total PL from NV$^\text{-}$ and NV$^0$ respectively. In the absence of PL from defects besides NV$^0$ and NV$^\text{-}$, $c_{\text{-}}+c_0=1$.  Post-analysis of our data shows little if any PL contributions in the 550-818 nm range from defects other than NV$^\text{-}$ and NV$^0$ for irradiation doses $\lesssim 1 \times 10^{19}$ e$^{\text{-}}/$cm$^2$.

\subsection{Photoluminescence spectrum decomposition}
\label{Sec:PhotoluminescenceSpectrumDecomposition}
To determine the charge state efficiency $\zeta$ using Eqn.~\eqref{eqn:chargestateefficiency}, the fractional contributions of each charge state in a given PL spectrum ($c_0, c_{\text{-}}$) are determined by fitting the acquired spectrum $I(\lambda)$ 
to a linear combination of NV$^\text{-}$ and NV$^0$ emission:  
\begin{equation}\label{eqn:pldecomposition}
I(\lambda) = M\Big[ c_{\text{-}}\hat{I}_{\text{-}}(\lambda)+c_0 \hat{I}_0(\lambda)\Big].
\end{equation}
Here, $M$ is a proportionality constant, and $\hat{I}_{\text{-}}(\lambda)$ and $\hat{I}_0(\lambda)$ denote area-normalized PL emission spectra for NV$^\text{-}$ and NV$^0$ respectively; the normalization satisfies $\int_\lambda \hat{I}_{\text{-}}(\lambda) d\lambda = 1$ and $\int_\lambda \hat{I}_0 (\lambda) d\lambda=1$, where the integral is taken over the wavelengths measured by the spectrometer. Note that, throughout this paper, the hat denotes dimensionless spectra which are normalized to unity area.

\begin{figure}[!b]
\begin{minipage}[t]{0.45\textwidth}
\vspace*{\fill}
\centering
        \includegraphics[width=\textwidth] {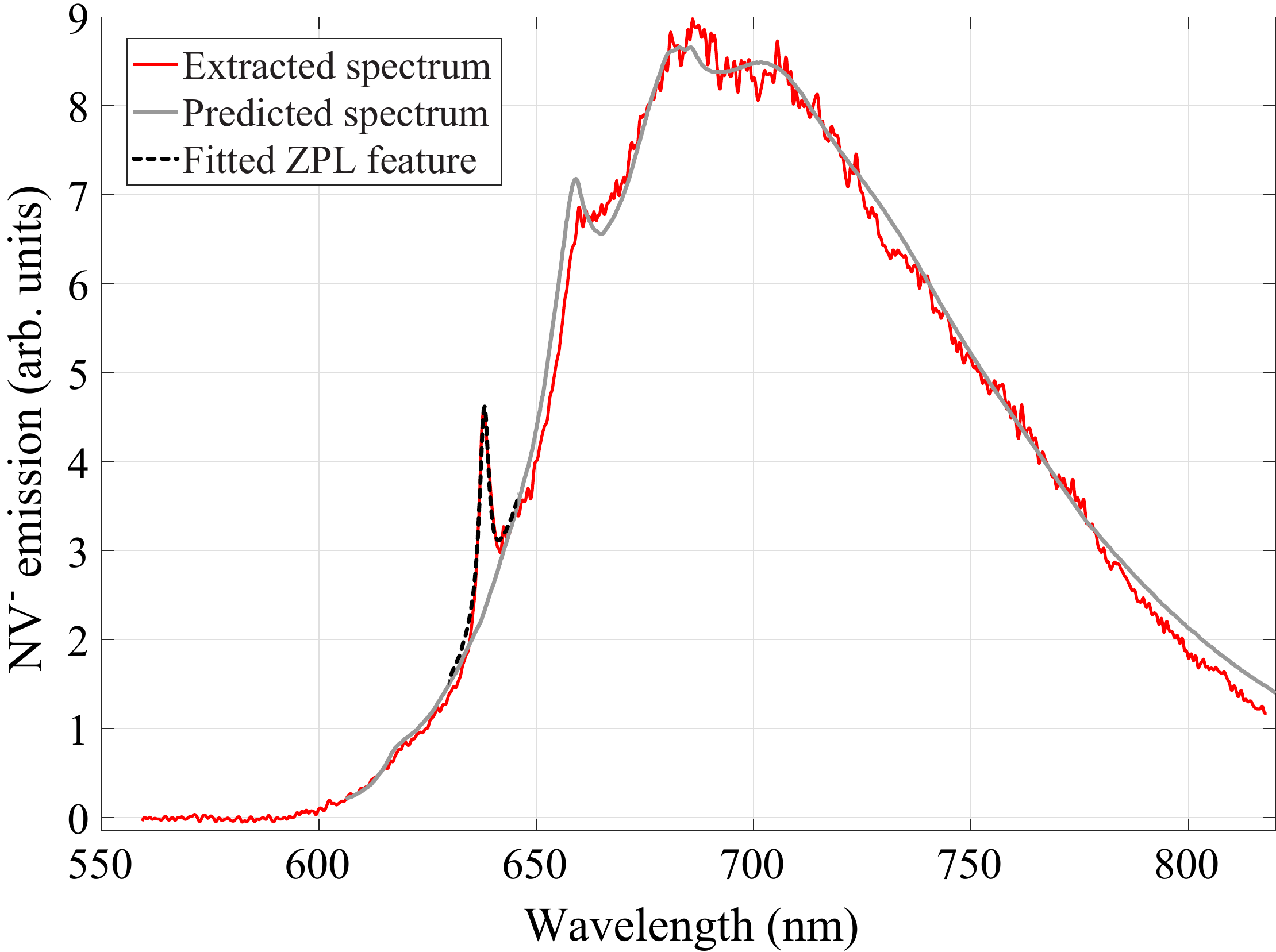}
         \par\vfill
        \end{minipage}
    \caption{\textbf{NV$^\mathbf{\text{-}}$ photoluminescence.} Extracted ({\color{red}{\rule[.6mm]{3mm}{.3mm}}}) and predicted ({\color{gray}{\rule[.6mm]{3mm}{.3mm}}}) NV$^\text{-}$ photoluminenscence emission spectrum at room temperature as described in the main text. The extracted and predicted curves are in good agreement. The ZPL feature itself ({\color{black}{\rule[.6mm]{1mm}{.3mm}}}\,{\color{black}{\rule[.5mm]{1mm}{.3mm}}}\,{\color{black}{\rule[.6mm]{1mm}{.3mm}}}) inserted here is not predicted a priori, but is consistent with the range of ZPL parameters found in the literature.}
\label{fig:NVminusTheoryVsExperiment}
\end{figure} 

\begin{figure}[!t]
\begin{minipage}[t]{0.45\textwidth}
\vspace*{\fill}
\centering
        \includegraphics[width=\textwidth] {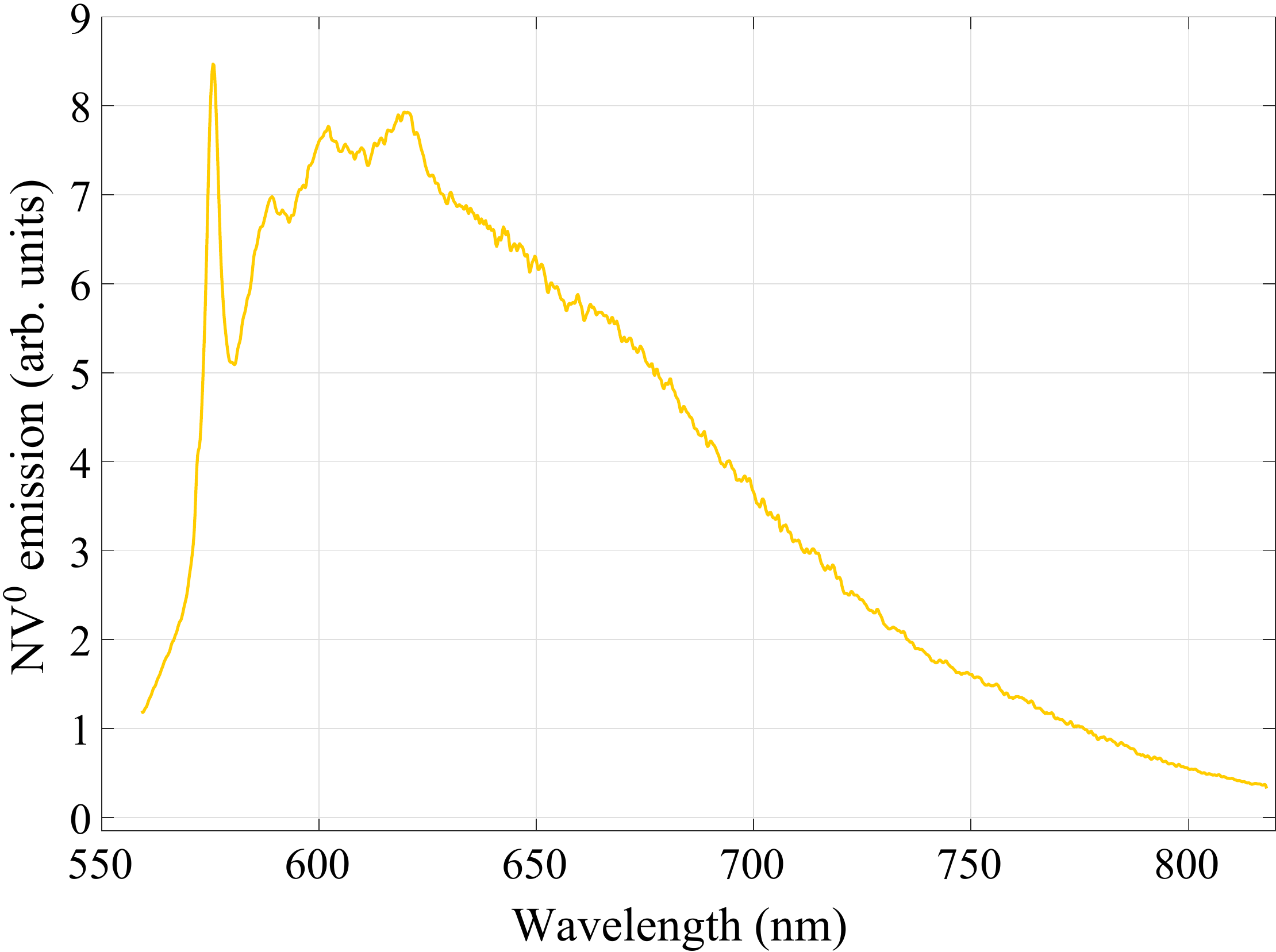}
         \par\vfill
        \end{minipage}
    \caption{\textbf{NV$^\mathbf{0}$ photoluminescence.} Extracted ({\color{dandelion}{\rule[.6mm]{3mm}{.3mm}}}) NV$^0$ photoluminescence emission spectrum at room temperature as described in the main text.}
\label{fig:NVzeroExperiment}
\end{figure} 

We now describe the process employed to determine $\hat{I}_0(\lambda)$ and $\hat{I}_{\text{-}}(\lambda)$, the accuracy of which are critical for the decomposition denoted in Eqn.~\eqref{eqn:pldecomposition} to be effective.  
PL spectra are recorded for all 28 diamonds using 18 different 532 nm  optical excitation intensities between 1 kW/cm$^2$ and 4 MW/cm$^2$ (see Appendix Sec.~\ref{Appendix:IntensityandSaturation} for intensity calculations). 
The PL spectra are area-normalized and sorted by the fractional PL contribution in the 560-600 nm band, which includes the NV$^0$ zero-phonon line (ZPL) at 575 nm. 
The area-normalized PL spectrum with the greatest 560-600 nm content is denoted $\hat{I}_0^\text{pre}(\lambda)$ and the area-normalized PL spectrum with the least 560-600 nm content is denoted $\hat{I}_{\text{-}}^\text{pre}(\lambda)$. 
The superscripted $\text{pre}$ label denotes the spectra are preliminary and require additional processing, as now  described. The spectra $\hat{I}_0^\text{pre}(\lambda)$ and $\hat{I}_{\text{-}}^\text{pre}(\lambda)$ are nominally dominated by PL emission from NV$^0$ and NV$^\text{-}$ respectively. 
Post-analysis investigation shows $\hat{I}_0^\text{pre}(\lambda)$ contains $\sim\!94\%$ NV$^0$ emission and $\sim\!6\%$ NV$^\text{-}$ emission, while  $\hat{I}_{\text{-}}^\text{pre}(\lambda)$ contains $\sim\!84\%$ NV$^\text{-}$ emission and $\sim\!16\%$ NV$^0$ emission. 
We assume that PL from other defects (SiV, V$^0$, etc.,) can be ignored; no ZPLs from such defects were observed for irradiation doses $\lesssim 10^{19}$ e$^{\text{-}}/$cm$^2$. 

The un-normalized NV$^\text{-}$ $I_{\text{-}}(\lambda)$ function is constructed using
\begin{equation}
I_{\text{-}}(\lambda) = \hat{I}_{\text{-}}^\text{pre}(\lambda)- \delta_0 \hat{I}_0^\text{pre}(\lambda),
\end{equation}
where the value of $\delta_0$ is adjusted so that $I_{\text{-}}(\lambda)$ exhibits no evidence of the NV$^0$ ZPL feature at 575 nm. The $I_{\text{-}}(\lambda)$ function is then area-normalized to yield $\hat{I}_{\text{-}}(\lambda)$. The $\hat{I}_{\text{-}}(\lambda)$ function is determined to be free of any NV$^0$ contribution to the 1$\%$ level or better (See Appendix Sec.~\ref{Appendix:BasisFunction}). The un-normalized NV$^0$ $\hat{I}_0(\lambda)$ function is then similarly constructed using
\begin{equation}
I_0(\lambda) = \hat{I}_0^\text{pre}(\lambda)- \delta_{\text{-}} \hat{I}_{\text{-}}(\lambda),
\end{equation}
where the value of $\delta_{\text{-}}$ is adjusted so the resulting $I_0(\lambda)$ exhibits no evidence of the NV$^\text{-}$ ZPL feature at 637 nm. This $I_0(\lambda)$ function is then area-normalized to yield $\hat{I}_0(\lambda)$.  Ensuring the $\hat{I}_0(\lambda)$ function is free of any NV$^\text{-}$ contribution is non-trivial; the NV$^\text{-}$ ZPL coincides with near-peak NV$^0$ vibronic sideband emission and is thus difficult to isolate and remove. Nevertheless, $\delta_{\text{-}}$ is determined to be $6.4\% \pm 3.3\%$, suggesting the resultant $\hat{I}_0(\lambda)$ function should exhibit at most a few percent contribution from NV$^\text{-}$. The extracted $\hat{I}_{\text{-}}(\lambda)$ and $\hat{I}_0(\lambda)$ functions \textcolor{black}{for bulk NV-diamond} are shown in Figs. \ref{fig:NVminusTheoryVsExperiment} and \ref{fig:NVzeroExperiment}, respectively.

As an independent check of the extracted $\hat{I}_{\text{-}}(\lambda)$ basis function, we adjust the low-temperature, one-phonon emission spectrum of the $^3$A$_2 \leftarrow ^3$E transition~\cite{Kehayias2013} for higher temperatures and multi-phonon transitions, as outlined in Ref.~\cite{Goldman2015}.
Specifically, the 300 K trace (Fig. 8) from Ref.~\cite{Goldman2015} is digitized~\cite{Engauge2018} and adjusted by the necessary $\omega^3$ weighting~\cite{Alkauskas2014} to yield an expected room temperature PL spectrum (less any ZPL contribution). The results are shown in Fig. \ref{fig:NVminusTheoryVsExperiment} and are in good agreement with our experimental determination of $\hat{I}_{\text{-}}(\lambda)$. As the spectrum derived from Ref.~\cite{Goldman2015} excludes the ZPL contribution, a comparison of the expected and measured data yields residuals whose main feature is the NV$^\text{-}$ ZPL; this residual data is well-fit by a Lorentzian with a Debye-Waller factor of 4.3 and a full width at half maximum (FWHM) linewidth of 2.6 nm. These values are well within the range of values observed in the literature \cite{acosta2009diamonds} (see Table \ref{HRTable}). 

\begin{figure}[t]
\begin{minipage}[t]{0.45\textwidth}
\centering
        \includegraphics[width=
        0.95\textwidth] {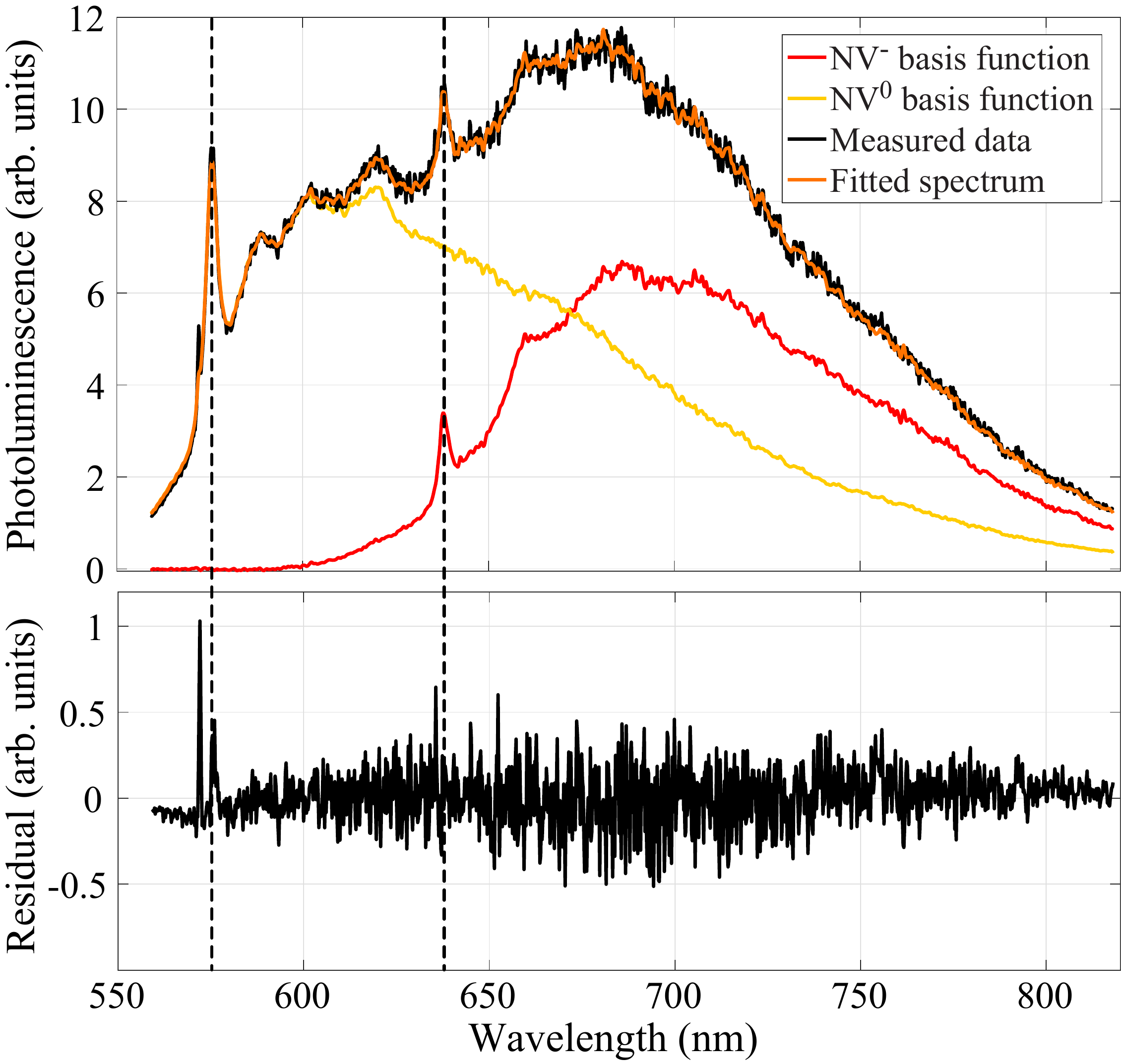}\vfill
        \end{minipage}
    \caption{\textbf{Example decomposition of diamond photoluminescence spectra.} a) PL spectra ({\color{black}{\rule[.6mm]{3mm}{.3mm}}}) for sample F, decomposed as a linear combination of the NV$^0$ ({\color{dandelion}{\rule[.6mm]{3mm}{.3mm}}}) and NV$^\text{-}$ ({\color{red}{\rule[.6mm]{3mm}{.3mm}}}) PL emission spectra.  The fit ({\color{orange}{\rule[.6mm]{3mm}{.3mm}}}) to the total PL is also shown. b) Residuals between the measured and fit data in a).  In both subfigures, the dashed black ({\color{black}{\rule[.6mm]{1mm}{.3mm}}} {\color{black}{\rule[.6mm]{1mm}{.3mm}}} {\color{black}{\rule[.6mm]{1mm}{.3mm}}}) vertical lines indicate the position of the the NV$^0$ and NV$^\text{-}$ ZPLs at 575 nm and 637 nm respectively.}
\label{fig:Basis_Function_Decomposition}
\end{figure}

\begin{figure}[b]
  \centering
        \includegraphics[width=0.45\textwidth]{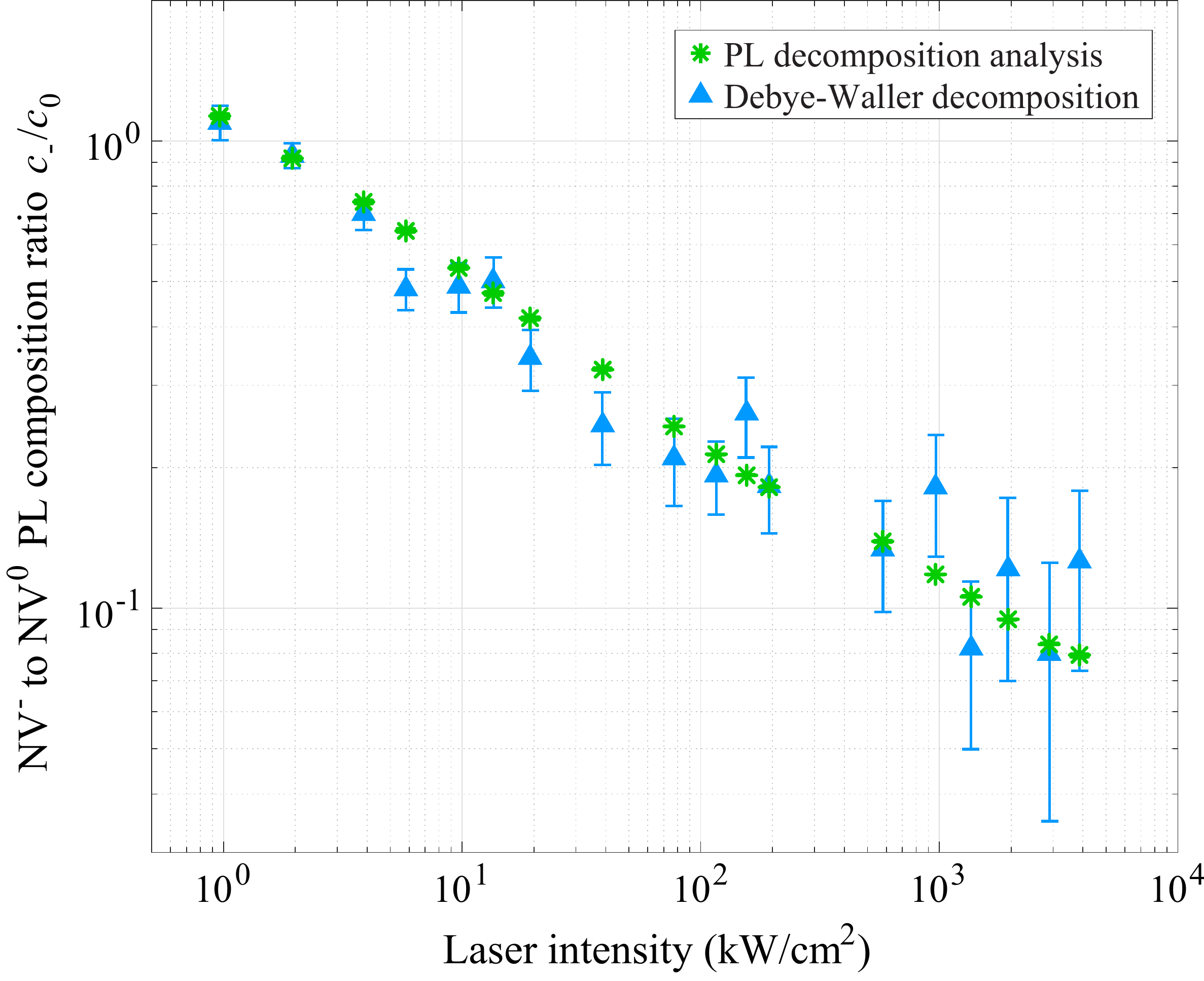}
            \caption{\textbf{Comparison of methods determining the PL composition ratio $\bm{c_{\text{-}}/c_0}$ for sample S.} The PL composition ratio $c_-/c_0$ is determined using both the photoluminescence decomposition analysis (\textcolor{springgreen}{\textbf{$\mathrlap{\times}+$}}) and the Debye-Waller decomposition (\textcolor{skyblue}{$\blacktriangle$}) for various laser intensities. Both methods analyzed the same data. The photoluminescence decomposition analysis produces $c_-/c_0$ ratios that are more consistent across laser intensities and exhibit lower error bars.}
        \label{fig:ZPLfitting}
\end{figure}

Using the determined $\hat{I}_{\text{-}}(\lambda)$ and  $\hat{I}_0(\lambda)$ basis functions, the PL spectra for all 28 diamonds at the 18 investigated optical excitation intensities are decomposed via Eqn. \ref{eqn:pldecomposition}. An example decomposition is shown in Fig. \ref{fig:Basis_Function_Decomposition}. To verify our approach, we compare the PDA method presented here to the Debye-Waller decomposition and observe that the PDA method is more robust to noise and produces more consistent results (Fig.~\ref{fig:ZPLfitting}). 
This improvement is likely due to the PDA method harnessing the entire $\sim\!200$ nm PL spectrum, compared to the Debye-Waller decomposition, which utilizes only data in the ZPL vicinity (a few nm). Further information on the Debye Waller decomposition is detailed in Appendix Sec.~\ref{Appendix:HuangRhysFactor}.  As an additional benefit, the PDA method does not require a priori knowledge of the Debye-Waller factors, which are observed to vary in the literature (see Table \ref{HRTable}). 

\subsection{Charge state photoluminescence ratio}

Accurate determination of the charge state efficiency $\zeta$ using Eqn. \ref{eqn:chargestateefficiency} requires first determining $\kappa_\text{532}$, the relative PL emission rate of NV$^\text{-}$ to NV$^0$ under low-intensity (i.e. well below saturation for either species) 532 nm optical excitation. Equivalently, $\kappa_{\lambda}$ can be formally defined as
\begin{equation}
   \kappa_{\lambda} = \Bigg[\frac{\int_{\lambda'} I_{\text{-}}(\lambda')d\lambda'}{\int_{\lambda'} I_0(\lambda')d\lambda'}\Bigg]_{[\text{NV}^0]=[\text{NV}^\text{-}]},
\end{equation}
where $\kappa_{\lambda}$ is evaluated in the low optical excitation intensity limit. 

We estimate the value of $\kappa_{532}$ using three independent methods.  First, the total integrated PL is proportional to
\begin{equation}\label{eqn:zetadeterminationscott}
\int_{\lambda} I(\lambda)\; d\lambda\; \propto\; \bigg([\text{NV}^0]+[\text{NV}^\text{-}]\kappa_{532} \bigg) \frac{2P}{\pi w_0^2},
\end{equation}
where $P$ and $w_0$ are the optical excitation power at the confocal volume and the $1/e^2$ intensity radius at the origin, respectively. The factor $2P/\pi w_0^2$ corresponds to the peak intensity of the exciting laser beam (see Appendix Sec.~\ref{Appendix:IntensityandSaturation}). The concentrations $[\text{NV}^0]$ and $[\text{NV}^\text{-}]$ are constrained at any given time to satisfy 
\begin{equation}
\label{NVconstraint}
    [\text{NV}^0]+[\text{NV}^\text{-}]=\text{constant}
\end{equation}
for a given sample. Note that Equation~\ref{NVconstraint}  implicitly assumes [NV$^\text{+}$] can be neglected. Even well below saturation intensity, the ratio [NV$^\text{-}$]/[NV$^0$] varies as a function of the applied 532 nm optical excitation  intensity, as shown in Fig.~\ref{fig:ZPLfitting}. Therefore, by varying the 532 nm optical excitation intensity to vary the ratio [NV$^\text{-}$]/[NV$^0$], the value of $\kappa_{532}$ can be determined. This method was employed for seven diamonds where $c_0 \approx c_{\text{-}}$ for 532 nm laser intensities $<$ 20 kW/cm$^2$ (i.e. well below saturation). Analyzing the combined data, we extract $\kappa_{532} = 2.5\pm0.5$; this result represents the first reported estimation of the relative PL emission rate of NV$^\text{-}$ to NV$^0$ and is one of the key results of this paper.

Second, the above-derived value for $\kappa_{532}$ is checked using third party data where the [NV$^\text{-}$]/[NV$^0$] ratio was varied via electrical control. The authors of Ref.~\cite{Kato2013} fabricate NV centers inside the insulating layer of a diamond-based PIN junction diode. By adjusting the bias voltage applied across the device, the authors directly controlled the [NV$^\text{-}$]/[NV$^0$] ratio (see Ref.~\cite{Kato2013} Fig. $\!$4). Using PL spectra taken for a variety of bias voltages (obtained directly from the authors of Ref.~\cite{Kato2013}), the charge state PL ratio $\kappa_{532}$ is estimated (see Appendix Sec.~\ref{Appendix:PLRatio}). This secondary method finds $\kappa_{532} = 2.17$. We note that this method implicitly assumes that the electric field used to vary the [NV$^\text{-}$]/[NV$^0$] ratio does not affect the dynamics of NV$^\text{-}$ or NV$^0$ beyond their charge state ratio. Both methods implicitly assume that [NV$^0$] + [NV$^\text{-}$] is conserved. 

Third, as a final check on the determined value of $\kappa_\text{532}$, we evaluated the [NV$^\text{-}$]/[NV$^0$] ratio for a test diamond not part of the rest of this study. UV-Vis measurements on this diamond suggest [NV$^\text{-}$]/[NV$^0$]$=\!1.33 \pm 0.2$. PL measurements at 1 kW/cm$^2$ and 2 kW/cm$^2$ intensity find the $c_{\text{-}}/c_0$ ratio to be = 2.6 and 2.1 respectively; a linear extrapolation (which we note is likely to underestimate $c_{\text{-}}/c_0$) to zero illumination suggests $c_{\text{-}}/c_0$= 3.1 in the dark. This analysis yields an estimate of $\kappa_\text{532} = 2.3 \pm 0.2$, although we note this linear extrapolation is likely to underestimate the actual value of $\kappa_{532}$. As all three methods are in good agreement, we estimate $\kappa_{532} = 2.5\pm0.5$.

The  determination of the relative florescence ratios of NV$^\text{-}$ to NV$^0$ is expected to aid in uncovering presently unknown dynamics of the NV$^0$, which is less well studied than NV$^\text{-}$. For example, this ratio may help determine the NV$^0$ $^2$\!A state stimulated emission cross section or the radiative quantum efficiency.

\section{Annealing and nitrogen conversion study}

For CVD-grown diamonds, the fraction of total nitrogen incorporated as NV centers during growth is typically quite low, with reported values ranging roughly from 0.0007 to 0.03 \cite{Hartlandthesis2014,Edmonds2012}. 
As discussed previously, this low as-grown NV conversion efficiency $\chi$ can be increased through electron irradiation and subsequent annealing. At the MeV-level energies typically employed in electron irradiation, NV creation via this mechanism depends significantly on irradiation dose, with behavior that can be simply described in the following three regimes:

\begin{enumerate}
\item  For irradiation doses which introduce substantially fewer vacancies than the grown-in native NV population, the observed quantity  [NV$^\text{T}$] is predominantly independent of the electron irradiation dose $D_{e}$.
\item At very high irradiation doses, the quantity [NV$^\text{T}$] saturates as every nitrogen is paired with a vacancy (to an order unity correction factor~\cite{Hartlandthesis2014}).
\item At certain intermediate irradiation doses, the value of [NV$^\text{T}$] is expected to vary linearly with the electron irradiation dose $D_{e}$~\cite{Hartlandthesis2014}. In this regime, NV centers created by irradiation and annealing outnumber the native NV centers, yet [V] $\ll$  [N$^\text{T}$] so a given additional vacancy is expected to create an additional NV.
\end{enumerate}

\begin{figure}[!b]
    \centering
        \includegraphics[width=0.45\textwidth]{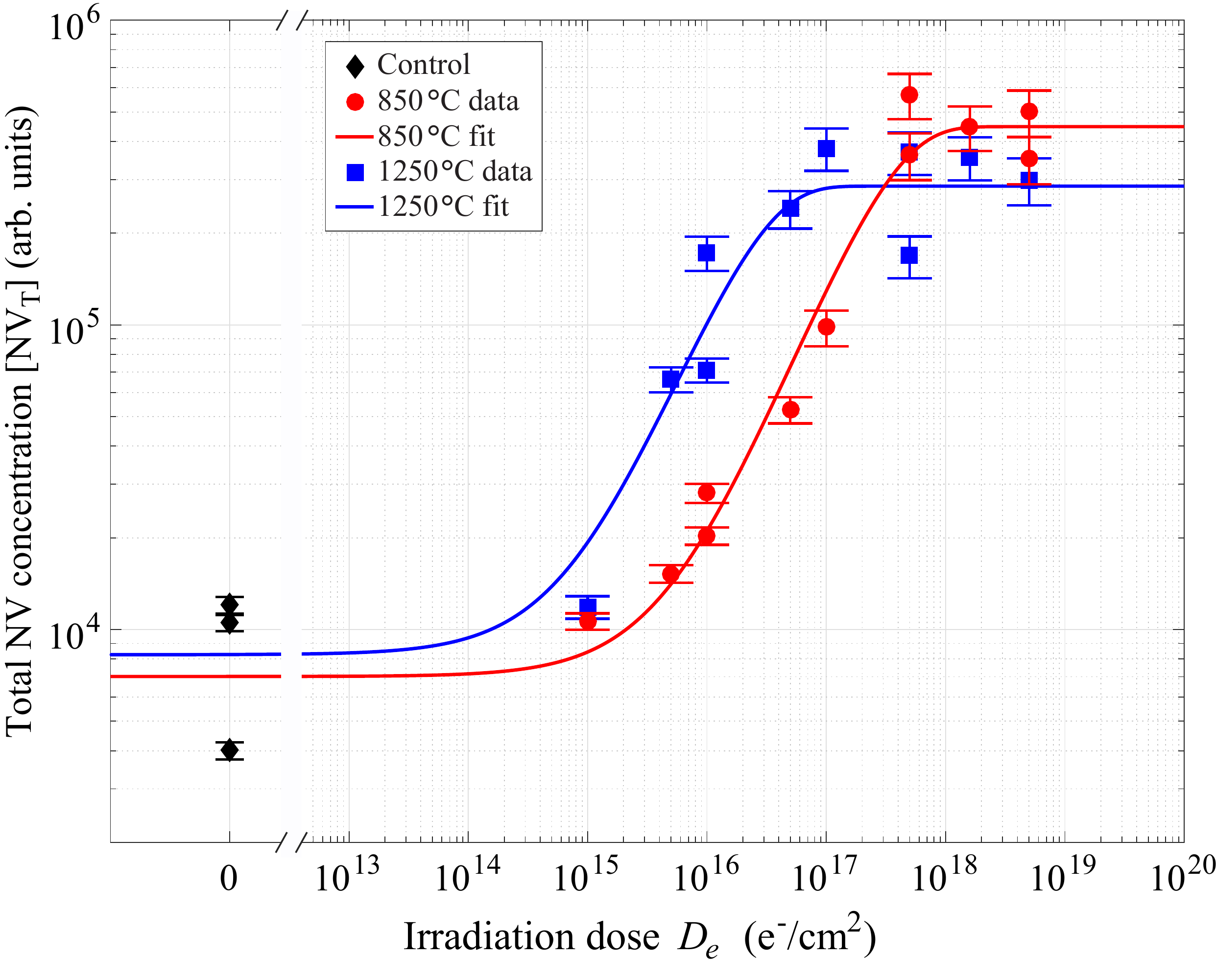}
        \caption{\textbf{Total NV concentration versus irradiation dose} \bm{$D_e$}\textbf{.} Data are shown for un-irradiated control diamonds ({\tiny$\blacklozenge$}), 850~$^\circ$C annealed diamonds (\textcolor{red}{\textbullet}), and 
        1250~$^\circ$C annealed diamonds (\textcolor{blue}{\tiny$\blacksquare$}). Error bars denote standard deviation of multiple measurements on the same sample.}
    \label{fig:Fluorversusdosage}
\end{figure}

We propose a simple phenomenological model to approximate NV annealing dynamics consistent with the above three limiting cases,
\begin{align}\label{eqn:plversusdosage}
[\text{NV}^\text{T}] &= [\text{NV}^\text{T}]_\text{gro}\\ \nonumber
&\;\;\;\;\; +\Big(F_c[\text{N}^\text{T}]-[\text{NV}^\text{T}]_\text{gro}\Big)\Big(1-e^{- \frac{B D_e}{F_c[\text{N}^\text{T}]-[\text{NV}^\text{T}]_\text{gro}}}\Big)
\end{align}
where $[\text{NV}^\text{T}]_{\text{gro}}$ is the total native NV concentration immediately following growth (i.e. before irradiation and annealing); the parameter $B$ is the vacancy creation rate per unit length; and the constant $F_c$ accounts for the empirical observation that $[\text{NV}^\text{T}]/[\text{N}^\text{T}]$ saturates at $0.6 \pm 0.07$~\cite{Hartlandthesis2014} in the limit of high irradiation dose. The following analysis therefore employs the estimate $F_c=0.6$ with the caveat that the true value of $F_c$ likely depends on the underlying diamond material and annealing conditions. For the diamonds in this study, [N$^\text{T}$] $\gtrsim 100\times [\text{NV}^\text{T}]_\text{gro}$; as a result, Eqn.~\eqref{eqn:plversusdosage} may be simplified to \cite{Edmonds2012}:
\begin{equation}\label{eqn:plversusdosage2}
[\text{NV}^\text{T}] \approx [\text{NV}^\text{T}]_\text{gro}+F_c[\text{N}^\text{T}]\Big(1-e^{- \frac{B D_e}{F_c[\text{N}^\text{T}]}}\Big).
\end{equation}

To test the validity of Eqn.~\eqref{eqn:plversusdosage2}, a correlate of the quantity [NV$^\text{T}$] is determined for each diamond as described herein. For each diamond, the PL spectrum is recorded under non-saturating 532 nm illumination ($\leq$ 20 kW/cm$^2$) and is decomposed into the NV$^\text{-}$ and NV$^0$ components (see Sec.~\ref{Sec:PhotoluminescenceSpectrumDecomposition}). Using the relative charge state PL emission ratio $\kappa_{532}$ at low excitation intensity and the fractional distribution between NV$^\text{-}$ and NV$^0$ in the PL ($c_{\text{0}}, c_{\text{-}}$), we adjust the PL contributions into a correlate of [NV$^\text{T}$]. That is, we set [NV$^\text{T}$] $=A(c_{\text{-}}+\kappa_{532}c_{\text{0}})$, where $A$ is proportional to the total PL. Such treatment is expected to be justified in the low intensity limit where the PL of NV$^\text{-}$ and NV$^0$ are both proportional to the optical excitation intensity (see Appendix Sec.~\ref{Appendix:IntensityandSaturation}). An example plot depicting [NV$^\text{T}$] for all 28 diamonds  is shown in Fig.~\ref{fig:Fluorversusdosage} for a 532 nm intensity of 6 kW/cm$^2$. 

The annealing data shown in Fig.~\ref{fig:Fluorversusdosage} illustrate that for intermediate irradiation doses $\sim 10^{17}$ e$^\text{-}$/cm$^2$, 
the 1250~$^\circ$C one-hour anneal was more effective at creating NV centers than the 850~$^\circ$C one-hour anneal. The results suggest that the 1 hour 850~$^\circ$C anneal was insufficient for most vacancies to diffuse to a substitutional nitrogen. In a simple model where non-nitrogen vacancy traps are ignored, we measure an approximate diffusion coefficient of $D = 1.8$ nm$^2$/s (see Appendix Sec.~\ref{Appendix:Diffusion} for additional details). 

Examination of the fit parameters in the phenomenological model given by Eqn.~\eqref{eqn:plversusdosage2} can give further information about the annealing and irradiation process. In particular, the data for samples annealed at 1250~$^\circ$C corresponds to a fitting parameter of $B = 0.52~\pm 0.26$ cm$^{-1}$ for [N$^\text{T}$]=118 ppb. This value of $B$ represents the vacancy creation rate for electron irradiation at 1 MeV when all recombination, including that which occurs during subsequent annealing, is accounted for. The value of $B$ determined here is approximate; if there are substantial vacancy sinks, the obtained value of $B$ will underestimate the true vacancy creation rate. However, conversely, the value of $B$ can then be used as a probe of the prevalence of vacancy sinks in diamond material of interest. Finally, we note that the value of $B$ obtained here falls within the range of values reported in the literature, 0.2 cm$^{-1}$ to 1.5 cm$^{-1}$ for electron energies between 1 and 5 MeV. Independent measurements using UV-Vis suggest a vacancy creation rate of $0.30\pm0.05$ cm$^{-1}$,  consistent with the measured value of $B$. See Appendix Sec.~\ref{Appendix:IrradiationStudy} for further discussion.

Additionally, Fig.~\ref{fig:Fluorversusdosage} shows that the fits for the 850~$^\circ$C and 1250~$^\circ$C anneals asymptote to different values of $[\text{NV}^\text{T}]$, with the former resulting in $\approx 50\%$ higher value than the latter. A comparison of means test yields a p-value of 0.05. Statistical significance aside, one plausible interpretation is that the increased mobility of one or more species at 1250~$^\circ$C is detrimental to the formation (or preservation) of NV centers. Another plausible hypothesis is that NV formation is less favorable at higher temperatures due to thermo-chromic charge transfer effects~\cite{Hartlandthesis2014}.

\section{$T_1$ and $T_2$ relaxation study}

\begin{figure}
    \centering
        \includegraphics[width=0.48\textwidth]{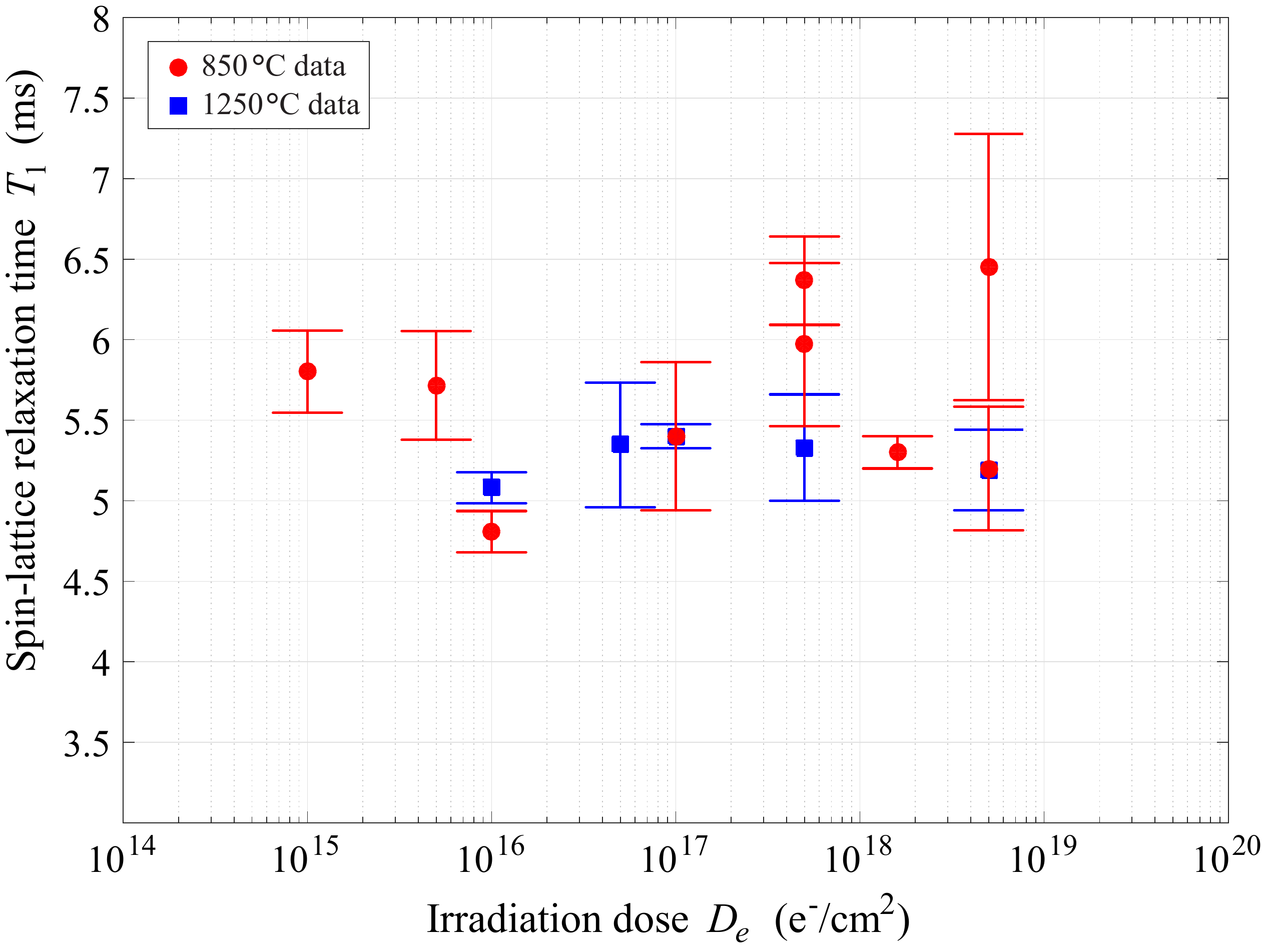}
        \caption{$\mathbf{T_1}$ \textbf{versus irradiation dose} $\bm{D_e}$\textbf{.} Data are shown for both 850~$^\circ$C annealed diamonds (\textcolor{red}{\textbullet}) and 1250~$^\circ$C annealed diamonds (\textcolor{blue}{\tiny$\blacksquare$}). The measured value of $T_1$ displays little if any dependence on either irradiation dose or anneal temperature for irradiation doses $D_e \! \leq  \! 5\times \! 10^{18}$~e$^\text{-}$/cm$^{2}$. Error bars denote standard deviation of multiple meausurements on the same sample.}
    \label{fig:T1vsDose}
\end{figure}

Following initialization into a given spin state, NV spins relax into the thermally mixed state from spin-lattice-induced interactions or other mechanisms~\cite{Choi2017}. The characteristic $T_1$ longitudinal relaxation time is particularly relevant for $T_1$ relaxometry measurements~\cite{tetienne2013spin,sushkov2014alloptical,hall2016detection}. The value of $T_1$ is evaluated for a randomly selected diamond subset (14 of the 28 samples) after irradiation and annealing. All 14 diamond samples exhibit values of $T_1$ between 4.8 and 6.5 ms.  For doses $\leq$ $5\times 10^{18}$ e$^\text{-}$/cm$^{2}$,  the data suggest $T_1$ exhibits little if any dependence on irradiation dose, as shown in Fig.~\ref{fig:T1vsDose}.

\begin{figure}[t]
    \centering
        \includegraphics[width=0.48\textwidth]{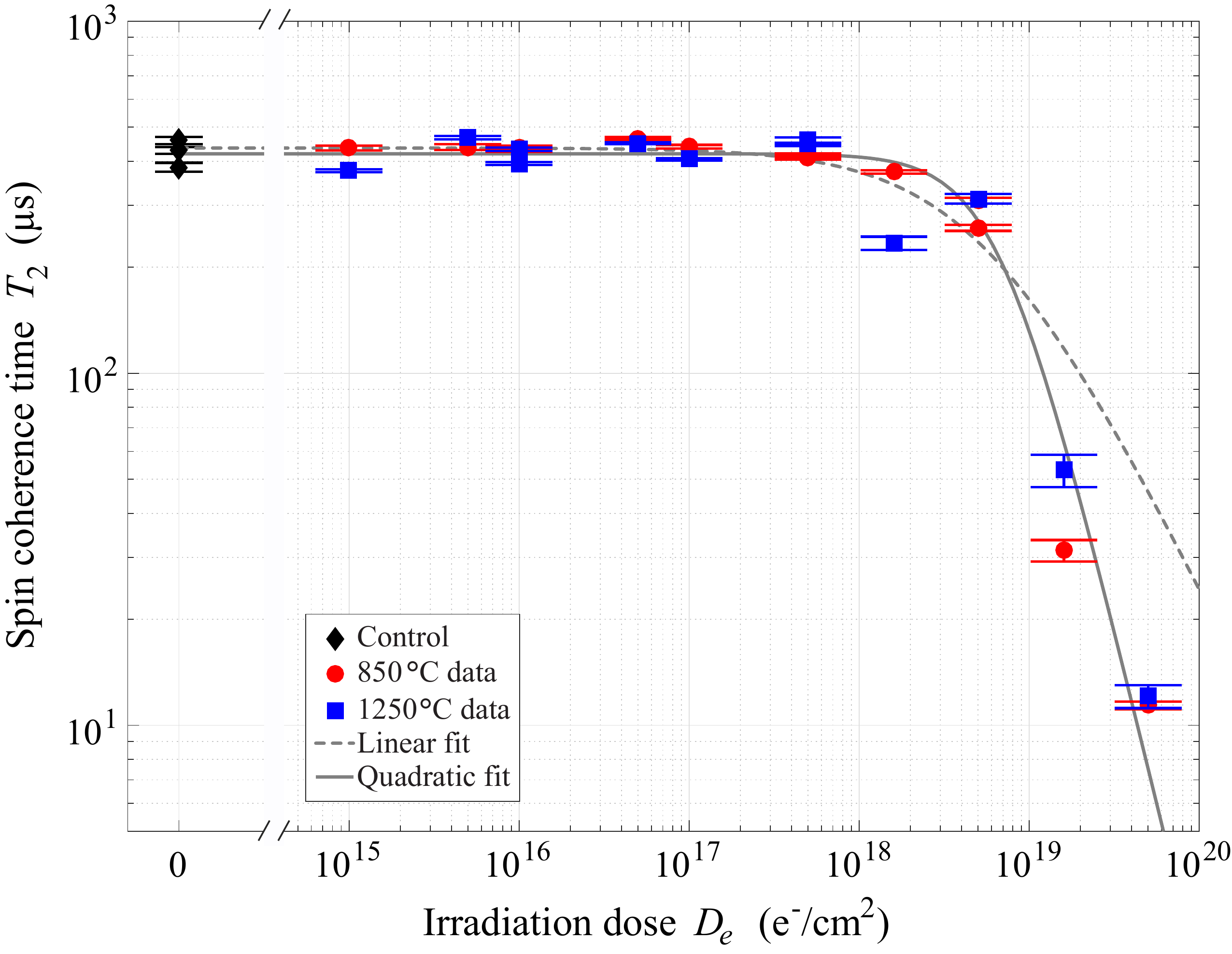}
        \caption{$\bm{T_2}$ \textbf{versus irradiation dose} $\bm{D_e}$\textbf{.} Data are shown for un-irradiated control diamonds ({\tiny$\blacklozenge$}), 850~$^\circ$C annealed diamonds (\textcolor{red}{\textbullet}), and 1250~$^\circ$C annealed diamonds (\textcolor{blue}{\tiny$\blacksquare$}). Little if any decrease in $T_2$ is observed for irradiation doses $D_e \lesssim 10^{18}$ e$^\text{-}$/cm$^{2}$. Data are fit to two models, one where $T_2$ decoherence is linearly proportional to dose ({\color{thegreen}{\rule[.6mm]{.65mm}{.5mm}}}\hspace{.35mm}{\color{thegreen}{\rule[.6mm]{.65mm}{.5mm}}}\hspace{.35mm}{\color{thegreen}{\rule[.6mm]{.65mm}{.5mm}}}) and another where $T_2$ decoherence is quadratic with dose ({\color{thegreen}{\rule[.6mm]{3mm}{.3mm}}}) as described in the main text. Error bars denote standard deviation of multiple measurements on the same sample.}
    \label{fig:T2vDose}
\end{figure}

The $T_2$ coherence time is an important parameter for AC magnetometry~\cite{Taylor2008,Pham2011},  quantum information~\cite{Dutt2007,Yao2012}, nanoscale NMR~\cite{mamin2013nanoscale,Devience2015,Glenn2018,Kehayias2017,Pfender2017nonvolatile,Aslam2017nanoscale,schmitt2017submillihertz,boss2017quantum}, single protein sensing~\cite{Lovchinsky2016}, and other applications involving NV-diamond quantum sensors. The $ T_2$ of a given species is determined in part by the surrounding paramagnetic spin bath~\cite{Bauch2018ultralong,Bauch2018b,Myers2017,Slichter1990}, and can be approximated as
\begin{equation}\label{eq:T2addingequation}
\frac{1}{T_2} \simeq \sum_{\text{X}}\frac{1}{T_{2,\text{NV}^\text{-}/\text{X}}}+\frac{1}{2 T_{1}},
\end{equation}
where $T_{2,\text{NV}^\text{-}/\text{X}}$ corresponds to the $T_2$ coherence time limited by NV$^\text{-}$ dipolar interactions with species X, $T_1$ is the longitudinal relaxation time~\cite{Jarmola2012}, and the summation is evaluated over all paramagnetic species X in the diamond. Similar to ion implantation\textcolor{red}{~\cite{Yamamoto2013,Oliveira2017,Iakoubovskii2002,kim2012electron}}, the electron irradiation process can introduce paramagnetic impurities into the diamond lattice~\cite{Lomer1973}, which degrade the $T_2$ coherence time. While annealing can partially mitigate this damage, empirical observations suggest some lattice damage will always remain~\cite{Lobaev2017,Balmer2009,E6patent2010WO149775,Oliveira2016,Oliveira2017}. Therefore, the irradiation doses at which $T_2$ and $T_1$ are compromised are quantities of interest for many applications employing NV ensembles.

The value of  $T_2$ is measured for all 28 diamonds using a Hahn-Echo sequence (See Appendix Sec.~\ref{Appendix:T2T1}) and is plotted versus irradiation dose in Fig.~\ref{fig:T2vDose}. To characterize the effects of the irradiation dose $D_e$ and annealing on $T_2$, the data is first fit to a linear model of the form $\frac{1}{T_2} = \frac{1}{W_1}\left[1+\frac{D_e}{W_2}\right]$.  We determine $W_1=436\pm 11$ $\upmu$s, which can be interpreted as the $T_2$ value in the absence of irradiation. However the poor fit of this \textit{linear model} at higher irradiation doses suggests the model does not capture the relevant dynamics. Instead, fitting the data to the \textit{quadratic model} $\frac{1}{T_2} = \frac{1}{W_1}\left[1+\frac{D_e^2}{W_3^2}\right]$ produces a better fit.  This fit does not improve with addition of a term linear in $D_e$. The best-fit parameters for this quadratic model are $W_1=420\pm10$ $\upmu$s and $W_3 =  6.8\pm0.8\times10^{18}$ cm$^{-2}$. The value of $W_3$ may be interpreted as the irradiation dose which decreases $T_2$ by $2\times$ relative to the un-irradiated value $T_2 = W_1$. 

Given that $W_3 \! \sim \! 400 \! \times \! \langle D_e^{\text{3dB}}\rangle$, where $\langle D_e^{\text{3dB}}\!\rangle=1.6\pm~0.8\!\times\! 10^{16}$ cm$^{-2}$ is the dose that converts half the initial nitrogen concentration to NV centers, the data suggest any desired NV conversion (i.e. $[\text{NV}]/[\text{N}^\text{T}]$) can be obtained with negligible decrease in $T_2$ from irradiation-induced defects. The failure of the linear model at the highest doses suggests the dominant source of decoherence observed at high doses is not the negatively charged vacancy. Further study is required to isolate the responsible decoherence mechanism.

\section{Summary and Outlook}

In this work, we presented a new room-temperature, all-optical technique for rapidly and quantitatively characterizing NV$^{\text{-}}$-to-NV$^0$ charge state ratio. In particular, we isolated the emission spectra of NV$^\text{-}$ and NV$^0$, allowing arbitrary diamond spectra to be decomposed into a linear combination of the NV$^\text{-}$ and NV$^0$ spectral contributions. This photoluminescence decomposition analysis exhibits better reproducibility and lower noise than the preexisting Debye-Waller-factor approach and does not require multiple excitation wavelengths~\cite{Fraczek2017}. Furthermore, to enable the \textit{quantitative} extraction of the [NV$^\text{-}$]/[NV$^0$] charge state ratio from these PL measurements, we also empirically determined for the first time the relative PL emission rate of NV$^\text{-}$ compared to that of NV$^0$, finding $\kappa_{532}=2.5\pm0.5$ under low-intensity 532-nm excitation. 

In addition to providing an expedient method to optimize irradiation and annealing of NV-diamond material targeting high-performance quantum sensing applications, the PDA technique and analyses presented are expected to help uncover unknown or poorly understood NV-diamond behavior. For example, the value $\kappa_{\lambda}$ can likely be used to place bounds on the branching ratios and quantum efficiency of the NV$^0$ $^2$E $\!\leftrightarrow\!$ $^{\!2\!}$A transition. Further, extensions of the PDA technique may be used to measure NV$^\text{-}$ and NV$^\text{0}$ saturation intensities, which are not precisely known and may vary for NV ensembles in different strain or charge environments where reported branching ratios differ~\cite{Robledo2011,Tetienne2012,Gupta2016}. Finally, extending the findings of this work---through additional study of vacancy creation, diffusion, and annealing as well as through further investigation of the mechanism for the NV$^\text{-}$ $T_2$ spin coherence time's quadratic dependence on irradiation dose---is expected to provide useful insight into the defect content of different diamond materials. Using these techniques and analyses to develop a more comprehensive understanding of diamond material composition and the effects of processing thereon is critical for advancing quantum sensing based on ensembles of NV centers in diamond.

We gratefully acknowledge the authors of Ref.~\cite{Kato2013}, Hiromitsu Kato, Marco Wolfer, Christoph Schreyvogel, Michael Kunzer, Wolfgang Müller-Sebert, Harald Obloh, Satoshi Yamasaki, and Christoph Nebel for providing raw data allowing us to confirm the measured value of $\kappa_\text{532}$. We thank Daniel Twitchen for insights into literature variation in irradiation outcomes.

\bibliography{thebib2016}

\appendix

\subsection{Diamond characteristics}
\subsubsection{Diamond nitrogen concentration estimation}
\label{Appendix:DiamondConcentration}

The primary contributors to NV$^\text{-}$ $T_2$ coherence times are $^{13}$C, residual neutral nitrogen, and spin-lattice interactions (i.e. $T_1$-related processes). Consequently, Eqn.~\eqref{eq:T2addingequation} can be used to estimate the value of [N$^\text{T}$].  The data in Fig.~\ref{fig:T2vDose} suggest $T_2 = 420 \pm 10 $ $\upmu$s for unirradiated diamonds in our study.  From Ref.~\cite{Bauch2018b} we expect $T_{2,\text{NV}^\text{-}/\text{N}} = 165 \pm 15$ ppm$\times \upmu$s\textcolor{black}{, where [N$^\text{T}$] is evaluated in ppm (i.e., a diamond with [N$^\text{T}$] = 1 ppm will have $T_{2,\text{NV}^\text{-}/\text{N}} = 165$ $\upmu$s, while a diamond with [N$^\text{T}$] = 2 ppm will have $T_{2,\text{NV}^\text{-}/\text{N}} = 165/2$ $\upmu$s $\approx 83$ $\upmu$s)}. For the $^{13}$C decoherence contribution, we use values reported in the literature for samples exhibiting natural isotopic abundance: Hall et al. theoretically predicted 780 $\upmu$s~\cite{Hall2014}; Mizuochi et al. measured 650 $\upmu$s for a single NV~\cite{Mizuochi2009}; Balasubranian et al.~\cite{Balasubramanian2009} measured 1.8 ms for a NV in 0.3$\%$ $^{13}$C diamond, which corresponds to a 504 $\upmu$s $T_2$ for natural abundance $^{13}\text{C}$; and Stanwix et al. reported 600 $\upmu$s for NV ensembles in high-purity diamonds with a natural abundance of $^{13}\text{C}$~\cite{Stanwix2010}.   Taking the average of these four values yields $T_{2,\text{NV}^\text{-}/^{13}\text{C}}=634 \pm 114$ $\upmu$s for diamonds with natural abundance $^{13}\text{C}$. We evaluate $T_1$ from Fig.~\ref{fig:T1vsDose} which yields $T_1 = 5.5 \pm 0.5$ ms and is in good agreement with data presented in Ref.~\cite{Jarmola2012}. Employing Eqn.~\eqref{eq:T2addingequation} with the above estimates suggests a mean nitrogen concentration [N$^\text{T}$] $=118 \pm 48$ ppb for the diamonds studied here.

This estimate is consistent with measurements in Ref.~\cite{Frangeskou2018} for standard-grade diamonds from Element Six. Test results of twenty diamonds of the same part number exhibit a mean nitrogen concentration of 120 ppb with a standard deviation of 16 ppb and minimum and maximum concentrations of 95 and 162 ppb respectively~\cite{Frangeskou2018}. Since the nitrogen concentration strongly affects CVD diamond growth rate, optical absorption, and other diamond properties, this parameter is carefully controlled during diamond growth, and it is perhaps unsurprising that the diamond samples employed in the work presented here exhibit similar nitrogen concentrations. 

\begin{table*}[ht]
\centering
	\begin{tabular}{ c  c  c  c  c }
		\hline
		\hline
            \;\; Sample \;\;&\;\; Pre-anneal, 1 hr \;\;&\;\; Irradiation dose at 1~MeV \;\;&\;\; First anneal, 1 hr \;\;&\;\; Second anneal, 1 hr \;\;\\ 
             & ($^{\circ}$C) & (e$^\text{-}$/cm$^2$) & ($^{\circ}$C) & ($^{\circ}$C)\\ 
             \hline
            \hline
            C1 & none & none & none & none \\ \hline 		
            C2 & none & none & none & none \\ \hline 		
            C3 & $850$ & none & $850$ & none \\ \hline 		
            A & $850$ & $1.0 \times 10^{15}$ & $850$ & none \\ \hline 		
            B & $850$ & $5.0 \times 10^{15}$ & $850$ & none \\ \hline 		
	        C & $850$ & $1.0 \times 10^{16}$ & $850$ & none \\ \hline 		
            D & $850$ & $1.0 \times 10^{16}$ & $850$ & none \\ \hline 		
            E & $850$ & $5.0 \times 10^{16}$ & $850$ & none \\ \hline 		
            F & $850$ & $1.0 \times 10^{17}$ & $850$ & none \\ \hline 		
            G & $850$ & $5.0 \times 10^{17}$ & $850$ & none \\ \hline 		
            H & $850$ & $5.0 \times 10^{17}$ & $850$ & none \\ \hline 		
            I & $850$ & $1.6 \times 10^{18}$ & $850$ & none \\ \hline 		
            J & $850$ & $5.0 \times 10^{18}$ & $850$ & none \\ \hline 		
            K & none & $5.0 \times 10^{18}$ & $850$ & none \\ \hline 		
           L & $850$ & $1.6 \times 10^{19}$ & $850$ & none \\ \hline 		
            M & $850$ & $5.0 \times 10^{19}$ & $850$ & none \\ \hline 		
            N & $850$ & $1.0 \times 10^{15}$ & $850$ &  $1250$ \\ \hline 		
            O & $850$ & $5.0 \times 10^{15}$ & $850$ &  $1250$ \\ \hline 		
            P & $850$ & $1.0 \times 10^{16}$ & $850$ &  $1250$ \\ \hline 		
            Q & $850$ & $1.0 \times 10^{16}$ & none &  $1250$ \\ \hline 		
            R & $850$ & $5.0 \times 10^{16}$ & none &  $1250$ \\ \hline 		
            S & $850$ & $1.0 \times 10^{17}$ & none &  $1250$ \\ \hline 		
            T & $850$ & $5.0 \times 10^{17}$ & none &  $1250$ \\ \hline 		
            U & $850$ & $5.0 \times 10^{17}$ & none &  $1250$ \\ \hline 		
            V & $850$ & $1.6 \times 10^{18}$ & $850$ &  $1250$ \\ \hline 		
            W & $850$ & $5.0 \times 10^{18}$ & none &  $1250$ \\ \hline 		
            X & $850$ & $1.6 \times 10^{19}$ & $850$ &  $1250$ \\ \hline 		
            Y & $850$ & $5.0 \times 10^{19}$ & $850$ &  $1250$ \\ \hline
            \hline
	\end{tabular}
\caption{\textbf{Electron irradiation and annealing parameters for diamonds used in this study.} Variables include an optional 850~$^\circ$C 1 hour anneal prior to irradiation, the 1 MeV electron irradiation dose $D_e$, a post-irradiation 850~$^\circ$C anneal for 1 hour, and an optional second anneal at 1250~$^\circ$C for 1 hour.}
\label{diamondtable}
\end{table*}

\subsubsection{Irradiation parameters}
\label{Appendix:DiamondIrradiation}

Diamonds were electron irradiated by a commercial supplier (Prism Gem, LLC). The electron energy of the Van de Graaff accelerator was fixed at 1 MeV. Doses of $1 \times 10^{15}$ e$^\text{-}$/cm$^2$ employ 2.5 mA beam current for 36 seconds.  Doses of $1 \times 10^{16}$ e$^\text{-}$/cm$^2$ and $1 \times 10^{17}$ e$^\text{-}$/cm$^2$ employ the same 2.5 mA beam current for 6 minutes and one hour respectively.  Finally, doses of $1 \times 10^{18}$ e$^\text{-}$/cm$^2$ and $1 \times 10^{19}$ e$^\text{-}$/cm$^2$ employ 25 mA of beam current for one hour and 10 hours respectively.  Irradiation doses are calculated by the formula $D_e = \frac{I t}{A q_e}$ where $I$ is the beam current, $A$ is the scan area over which the electron beam is rastered, $q_e$ is the electron charge, and $t$ is the irradiation time. Table~\ref{diamondtable} details irradiation doses for the 28 diamonds used in this study. 

\subsection{Experimental details}
\subsubsection{Confocal microscope}
\label{Appendix:ConfocalMicroscope}

\begin{figure*}
  \centering
      \includegraphics[width=0.85\textwidth]{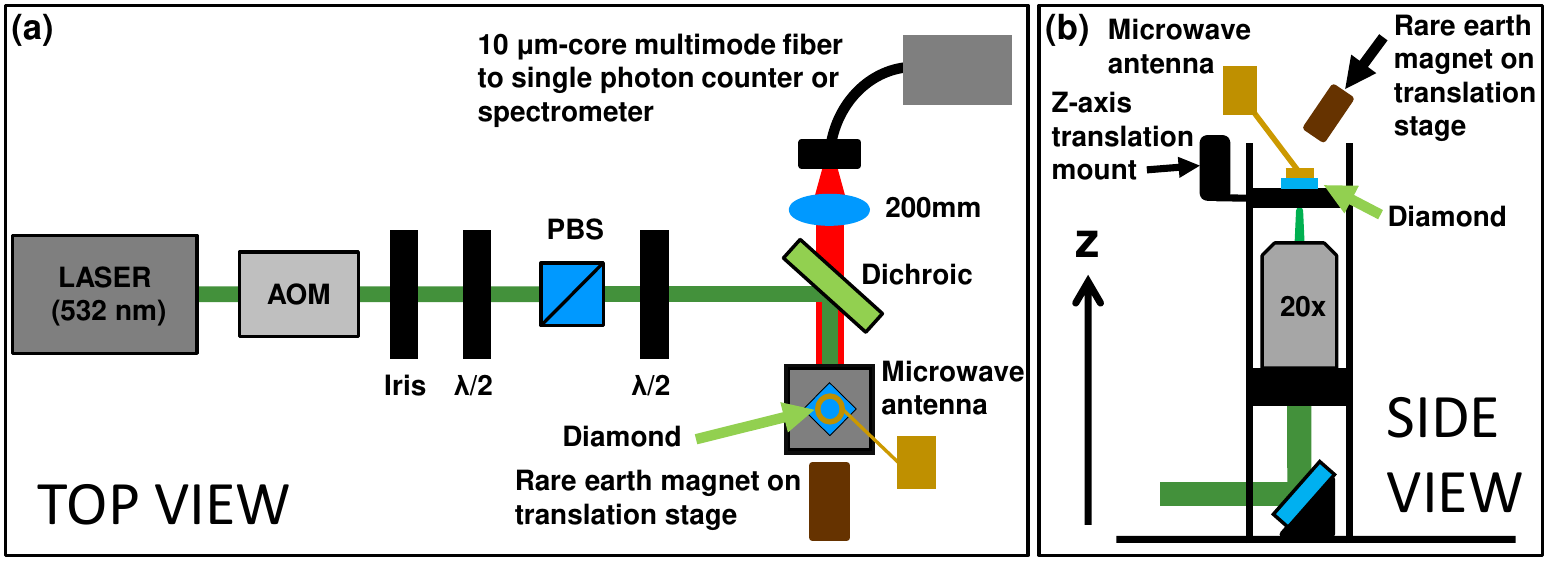}  
       \caption{\textbf{Overview of experimental setup.} (a) Top-down schematic of the confocal microscope detailing excitation and collection optics. The optical system is designed so that the numerical aperture of the collection (red) is greater than the numerical aperture of the excitation (green), allowing PL collection from only the center of the excitation volume. (b) Vertical view of confocal microscope in proximity to the diamond.}
        \label{simplifiedsetup}
\end{figure*}

A home-built confocal microscope, shown diagrammatically in Fig.~\ref{simplifiedsetup}, allows PL spectroscopy and coherence measurements of the diamond samples.
Green 532 nm laser light (Coherent Verdi G5) is sent through an acousto-optic modulator (Crystal Technology/Gooch $\&$ Housego 3200-147). The zeroth order beam is discarded while the first order beam  passes through a half-wave plate, reflects off a dichroic mirror, and is focused onto the diamond with a commercial microscope objective (Mitutoyo M Plan Apo 20X 378-804-3) with 0.42 NA, a $f_\text{obj}=10$ mm focal length, and a $D_\text{ba}=8.4$ mm back aperture. A permanent magnet mounted on a translation stage creates a static bias magnetic field \textcolor{black}{of moderate strength 40-60 gauss} aligned along one of the four diamond crystallographic directions, allowing individual addressing of the magnetic resonances associated with the NV sub-ensemble oriented along the bias field. This bias field alignment prevents $T_2$ measurements from being skewed by the effects of $^{\text{13}}$C.   

The diamond is mounted over a hole in a 170-$\upmu$m-thick glass coverslip attached to a stainless steel ring (Thorlabs, RS1M). Microwaves are applied to the diamond from above using a loop antenna. For $T_2$ and $T_1$  measurements, the interrogation volume is located at the top of diamond, closest to the microwave loop to achieve maximum MW drive strength. For PL measurements, the interrogation volume is located near the bottom diamond surface to minimize spherical aberration effects from the diamond's $n_d\approx2.41$ refractive index.

Light emitted from the NV-ensemble is collected by the 20X objective and subsequently coupled using a 200 mm focal length lens into 10-$\upmu$m-core multi-mode fiber (Thorlabs, M64L02) that serves as a pinhole.  This fiber is connected to either a spectrometer (Princeton Instruments, IsoPlane SCT 320) for photoluminescence measurements or to a fiber connected to an avalanche photodiode single photon counter (Excelitas SPCM-AQRH-FC). The spectrometer is both wavelength and intensity calibrated (Princeton Instruments Intelical). The intensity calibration removes instrumental artifacts in the PL spectra that arise from the differing spectral responses from various optical elements in the spectrometer (mirrors, diffraction grating, CCD camera, etc.). Counts registered by the single photon counter are recorded by a DAQ (National Instruments, PCIe-6251), which is in turn gated by a pulse generator (SpinCore Pulse Blaster ESR-PRO-500). The pulse generator is used to gate the optical excitation, microwave radiation, and single photon counter measurement.
\begin{figure}
  \centering
        \includegraphics[width=0.45\textwidth]{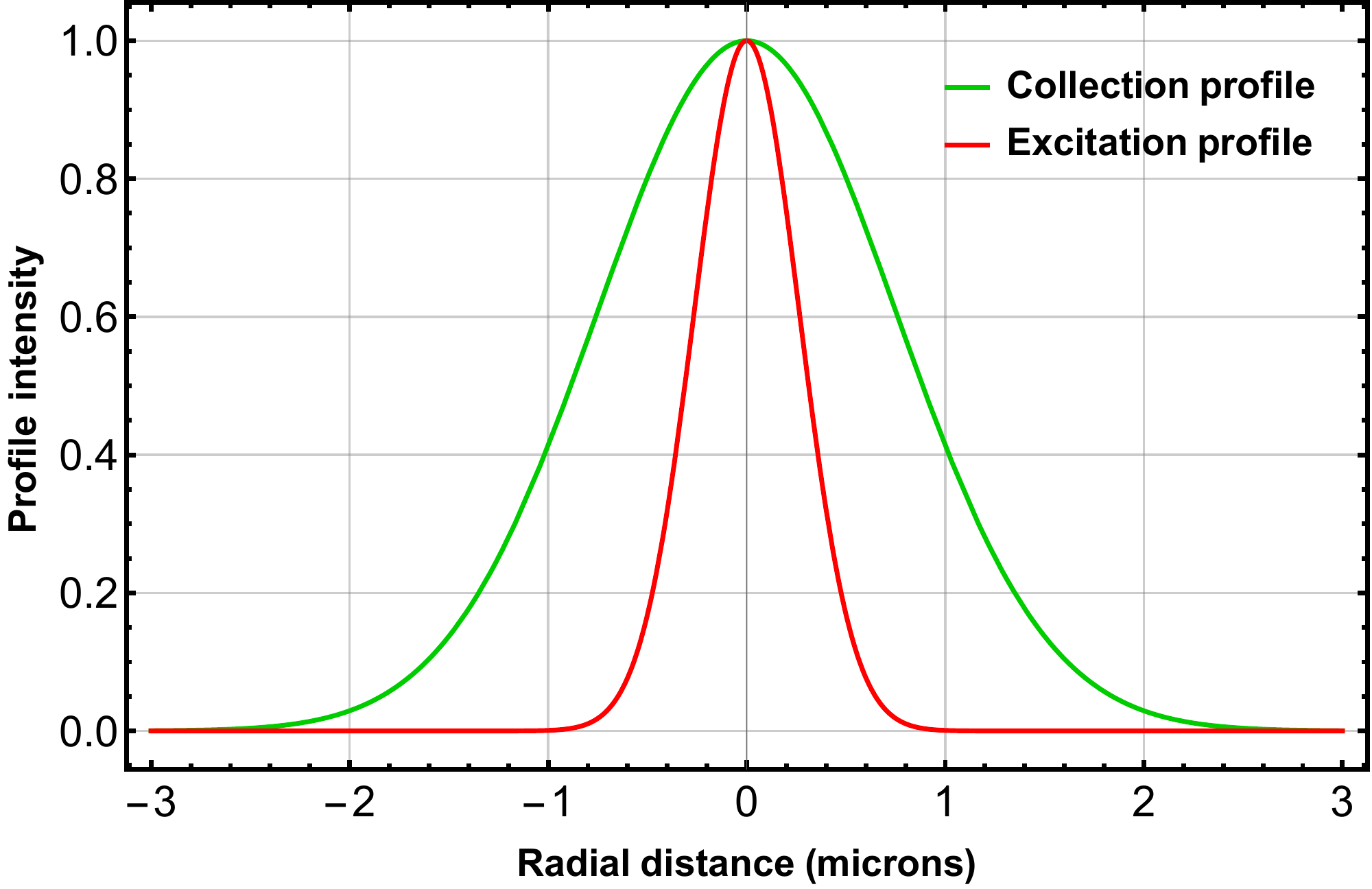}
            \caption{\textbf{Radial excitation and collection profiles at the center of the confocal volume.} The collection profile for 700 nm light is made smaller than the 532 nm excitation profile to allow PL collection only from NVs with near-uniform optical illumination intensity.}
        \label{excitationcollectionprofiles}
\end{figure}

\subsubsection{Intensity calculations and saturation}
\label{Appendix:IntensityandSaturation}
To calculate the intensity associated with a given excitation laser power, we model the intensity of the green laser beam focused by the objective as a Gaussian profile:
\begin{equation}
I(r,z) = \frac{2P}{\pi w(z)^2} \exp{\left(-\frac{2r^2}{w(z)^2}\right)},
\end{equation}
where $P$ is the incident 532 nm power corrected for any transmission losses, $w(z) = w_0 \sqrt{1+(z/z_R)^2}$ is the $1/e^2$ intensity radius, $z_R = \pi n_{d} w_0^2/\lambda$ is the Rayleigh range (adjusted for the diamond index of refraction $n_{d}$), $r$ is the radial distance in the plane transverse to the beam, $\lambda$ is the wavelength in free space, and $z$ is the axial distance from the origin, defined as the center of the collection volume~\cite{SalehTeich}. The $1/e^2$ intensity diameter at the origin is $2w_0$. The peak intensity $I_0$ is
\begin{equation}
I_0 = \frac{2P}{\pi w_0^2}.   
\end{equation}
A beam profiler (Newport LBP2-HR-VIS) measures the spot size of the Gaussian laser beam right before the objective to be about 2.25 mm in diameter ($1/e^2$). For an input diameter $D_{in}=2.25$ mm, laser wavelength $\lambda = 532$ nm, and objective focal length $f_{obj} = 10$ mm, the focused excitation beam waist is $w_0^{\text{ex}} = \frac{2\lambda f_{obj}}{\pi D_{in}}=$ 1.5 $\upmu$m~\cite{SalehTeich}. This yields a Rayleigh range $z_R = 32$ $\upmu$m for the excitation beam.

The saturation intensity of the NV$^\text{-}$ center is on the order of 100 kW/cm$^2$~\cite{Wolf2015} with the saturation intensity of the NV$^0$ likely to be higher. For measurements relating collected PL to [NV$^\text{T}$], excitation intensities of 3.3 kW/cm$^2$ or less were employed.

\subsubsection{Excitation and collection volume}
\label{Appendix:ExcitationCollectionVolume}
Interpretation of photoluminescence and related derived data is most straightforward if the interrogated NVs receive near-uniform optical excitation intensity. To ensure that the NVs in the collection volume are near-uniformly illuminated with 532 nm light, the collection volume is restricted to the central region of the excitation volume.  In contrast to the 532 nm excitation, the collected fluorescence makes full use of the objective 0.42 numerical aperture. As the objective back aperture $D_\text{ba}$ is 8.4 mm, the collection beam waist is $w_0^{\text{col}} = \frac{2\lambda f_\text{obj}}{\pi D_\text{ba}}$, which is $w_0^{\text{col}} = 0.53 \;\upmu$m for 700 nm light. A radial slice of the excitation and collection profiles is shown in Fig.~\ref{excitationcollectionprofiles}.

\subsection{Charge state photoluminescence details}
\subsubsection{Photoluminescence measurement protocol}
PL spectroscopy measurements are performed with a spectrometer and an attached CCD camera (Princeton Instruments, PIXIS 100). A half wave plate tunes the incident light polarization so that all four NV orientations are excited equally. To avoid signal to noise degradation by modal noise (i.e. speckle) in the multi-mode fiber delivering light to the spectrometer, the fiber is taped to a speaker-head driven by a 200 Hz sinusoidal voltage~\cite{Baudrand2001}.  Multiple PL measurements are averaged together and normalized by the exposure time. \textcolor{black}{The homogeneity of the commercial diamonds employed in this study was characterized by measuring PL spectra in 5 different spots in each of a subset of 8 diamonds; the PL intensity and extracted NV charge state ratios were found to vary by a maximum of 10\% across each diamond sample, indicating reasonable homogeneity within each diamond.}

\subsubsection{The Huang-Rhys and Debye-Waller factors}\label{Appendix:HuangRhysFactor}

A commonly used method to approximate the charge state efficiency is the Debye-Waller (DW) factor $\tilde{S}$, which is defined in terms of the relative weight of the ZPL fluorescence:
\begin{equation}
\tilde{S}_{\text{X}} = -\ln \left(\frac{I_\text{ZPL,\text{X}}}{I_\text{total,\text{X}}}\right),
\end{equation}
where $I_\text{ZPL,\text{X}}$ is the integrated fluorescence of the zero-phonon line of NV charge state X's PL spectra and $I_\text{total,\text{X}}$ is the total integrated fluorescence of that charge state~\cite{acosta2009diamonds}. In the limit where the ZPL intensities can be measured with sufficient accuracy, the DW factor would enable accurate determination of the relative fluorescence ratio of NV$^\text{-}$ and NV$^\text{0}$ for a given sample. However, the relative fluorescence ratio alone is insufficient to accurately determine the charge state efficiency, due to different dynamics (lifetimes, absorption cross sections, etc.,) of the NV$^\text{-}$ and NV$^\text{0}$ systems. The Debye-Waller factor can be derived from a theoretical treatment of the electron-phonon coupling for defects in the diamond lattice by approximating the coupling to be linear in the nuclear displacement~\cite{Maradudin1966}.  

\begin{table}
\centering
	\begin{tabular}{ c  c  c  c }
		\hline
		\hline
         NV$^\text{-}$ Debye-Waller  & Temp. (K) & Diamond & Ref. \\
            Factor $\tilde{S}$ &  & type &  \\
            \hline
            \hline
            2.65 & $\sim$4 & Ensemble, Ib & \cite{Davies1974} \\ \hline 		
            4.7 & 300 & Ensemble, Ib & \cite{Davies1974} \\ \hline 		
            4.79, 4.25, 2.79 & Room temp. & Single NVs & \cite{Kilin2000}\\ \hline 		
            3.9, 4.0 & Room temp. & Ensemble, Ib & \cite{acosta2009diamonds} \\ \hline 		
	        3.49 & 4 & Ensemble, Ib & \cite{Kehayias2013} \\ \hline 		
            3.45 & 8 & Ensemble, Ib & \cite{Alkauskas2014} \\ \hline	\hline	
	\end{tabular}
\caption{\textbf{Debye-Waller factors for NV$^\text{-}$ as reported in the literature.}}
\label{HRTable}
\end{table}

Frequently in the literature the Debye-Waller factor is confused with the Huang-Rhys (HR) factor. In Ref.~\cite{HuangRhys1950}, Huang and Rhys defined the Huang-Rhys factor to be the average number of photons of a $k$-phonon transition emitted in an optical transition~\cite{Alkauskas2014}. Here the ZPL is weighted with an additional pre-factor of the cube of the optical emission frequency, which decreases the value of the DW factor compared to the HR factor.  To avoid confusion, we follow the naming convention suggested by Ref.~\cite{Alkauskas2014}, denoting the Huang-Rhys factor by $S$ and the Debye-Waller factor by $\tilde{S}$.

During this study, we found determination of the charge state efficiency using the Debye-Waller decomposition to be non-ideal, since PL features from other defects can obscure the ZPL, making accurate determination of the DW (or HR) factor difficult. This is most problematic for the NV$^\text{-}$ ZPL, which lies within the NV$^0$ phonon sideband emission spectra. Additionally, the ZPL prominence relative to total PL is small at room temperature for both NV$^\text{-}$ and NV$^0$.  Due in part to both non-idealities, there exists substantial variation in the DW factors reported in the literature.  Measured values of $\tilde{S}$ for NV$^\text{-}$ in the literature are shown in Table~\ref{HRTable}.  The DW factor for NV$^0$ is less-studied and reported to be $\tilde{S} = 3.3$ at low temperature~\cite{Zaitsev2001}. 

To calculate the fractional contributions of NV$^0$ and NV$^\text{-}$ to total PL using the Debye-Waller factor, we fit and subtract a local linear background to the ZPL of each charge state and fit a Gaussian to the modified ZPL.  The ZPL areas $I_\text{ZPL,0}$ and $I_\text{ZPL,-}$ for NV$^0$ and NV$^\text{-}$ respectively are weighted with the DW factors reported in Ref.~\cite{acosta2009diamonds}: $\tilde{S}_0 = 3.3$ for NV$^0$ and $\tilde{S}_{\text{-}}=4.3$ for NV$^\text{-}$.  Using the form of $\zeta$ in Eq.~\ref{eqn:chargestateefficiency} with the weighted ZPL areas (normalized by the charge state PL emission rate $\kappa_{532}$) gives the charge state efficiency as determined with the Debye-Waller decomposition, $\zeta^{\text{DW}}$:
\begin{equation}
\zeta^{\text{DW}} = \frac{I_\text{ZPL,-}e^{\tilde{S}_{\text{-}}}}{\kappa_{532}I_\text{ZPL,0}e^{\tilde{S}_0}+I_\text{ZPL,-}e^{\tilde{S}_{\text{-}}}}.
\end{equation}

The Debye-Waller decomposition provides fair performance for diamonds with low NV$^0$ fluorescence, which allow for easy extraction of the NV$^\text{-}$ ZPL area~\cite{acosta2009diamonds}.  However, for diamonds with a high fraction of NV$^0$ fluorescence (in this study, diamonds with higher irradiation doses), determining the NV$^\text{-}$ ZPL area becomes difficult. Figure~\ref{fig:ZPLfitting} shows the PL ratio (i.e. $c_{\text{-}}/c_0$) determined with the Debye-Waller decomposition for sample S for various laser excitation intensities. The increasingly large error bars and systematic errors at higher laser intensities reflects the difficulty of determining the NV$^\text{-}$ ZPL area when the NV$^0$ photoluminescence dominates. Not only does the basis function decomposition lead to lower statistical errors when determining the charge state fluorescence ratio, but the method is particularly advantageous at the higher laser intensities where the Debye-Waller decomposition becomes unreliable.

\subsubsection{Construction of basis functions}
\label{Appendix:BasisFunction}

As the NV$^0$ ZPL lies outside the NV$^\text{-}$ PL spectrum, the NV$^\text{-}$ basis function can be created nearly free from NV$^0$. The little if any impurity is bounded in two ways and the basis function shape can be checked itself. First, by comparing peak deviations of the NV$^\text{-}$ basis function around 575 nm, it is determined that the NV$^\text{-}$ basis function contains $<0.015$ NV$^0$. Second, fitting a Lorentzian to the vicinity of 575 nm with FWHM bounded to lie in the range [2,5] nm suggests at most the NV$^\text{-}$ basis function contains 0.003 NV$^0$. Third, the extracted NV$^\text{-}$ basis function is in good agreement with theoretical predictions~\cite{Goldman2015,Kehayias2013,Alkauskas2014}, derived from experimentally determined single phonon sideband spectra from the $^3\!\text{A}_2\! \leftrightarrow ^3$E transition (see main text Fig.~\ref{fig:NVminusTheoryVsExperiment}). Given the good fit with theory and the lack of any discernable NV$^0$ ZPL in the NV$^\text{-}$ basis functions, we conclude that any contamination of the NV$^\text{-}$ basis function with NV$^0$ is on the 1$\%$ level or below. Residual fits to the $\hat{I}_0^{\text{pre}}$ function suggest $0.064 \pm 0.033$ is NV$^\text{-}$. As this component is removed, we expect the any residual component of NV$^\text{-}$ in the NV$^0$ basis function is estimated to occur at the $3\%$ level or less.

\textcolor{black}{Further note that the process for constructing basis functions described in the main text is robust to the choice of initial spectra for $\hat{I}_0^{\text{pre}}(\lambda)$ and $\hat{I}_{\text{-}}^{\text{pre}}(\lambda)$. For example, repeating the basis function construction process with different yet comparable spectra to those employed in the main text yields a maximum 3\% difference in the extracted basis functions and a maximum 3\% difference in the PL decomposition results from a subset of 8 samples. However, due to the difficulty in fully isolating and removing NV$^{\text{-}}$ residual contribution from the NV$^0$ basis function, selecting initial spectra with maximal NV$^0$ and NV$^{\text{-}}$ contributions for the $\hat{I}_0^{\text{pre}}(\lambda)$ and $\hat{I}_{\text{-}}^{\text{pre}}(\lambda)$ functions, respectively, is recommended.}


\subsubsection{Secondary determination of $\kappa_{532}$}
\label{Appendix:PLRatio}
To obtain the total NV content for the irradiation model (Eqn.~\eqref{eqn:plversusdosage2}) or to determine the charge state efficiency, we need to correct for the relative PL emission rate of NV$^\text{-}$ to NV$^0$. In the main text, we find $\kappa_{532} = 2.5\pm 0.5$ by varying the [NV$^\text{-}$]/[NV$^0$] ratio using different well-below-saturation 532 nm optical excitation intensities. To independently check the measured value of $\kappa_{532}$, we exploit 3rd party data, where the [NV$^\text{-}$]/[NV$^0$] ratio is varied electrically. Specifically, we employ the data from Ref.~\cite{Kato2013} Fig. 4a where a bias voltage is used to vary the [NV$^\text{-}$]/[NV$^0$] ratio. More positive bias voltages skew the favored charge state towards NV$^0$ while more negative bias voltages skew the favored charge state towards NV$^\text{-}$. 

We denote the PL spectra with 0 Volts applied as $f'_{\text{NV}^\text{-}}(\lambda)$ and the PL spectra with 35 Volts applied as $f'_{\text{NV}^0}(\lambda)$, which correspond to when the diamond under study in Kato et al. is primarily in the NV$^\text{-}$ and NV$^0$ charge states respectively. Note that these spectra are not area-normalized. From the raw data for Ref.~\cite{Kato2013} Fig. 4a provided by the authors, we calculate a \textit{preliminary} charge state PL ratio 
\begin{equation}\label{kappaprimeone}
\kappa'_{532} = \bigg[ \frac{\int_{\lambda} f'_{\text{NV}^\text{-}}(\lambda) d\lambda}{\int_{\lambda} f'_{\text{NV}^0}(\lambda) d\lambda} \bigg]_{[\text{NV}^0]=[\text{NV}^\text{-}]}.
\end{equation}
The above value of $\kappa_{532}'$ will need to be corrected since the voltage only partially influences the charge state. Straightforward evaluation gives $\kappa_{532}' = 1.687$.

We now describe how $\kappa_{532}'$ is corrected to obtain the true value of $\kappa_{532}$.
We desire to obtain the true charge state PL ratio $\kappa_{532} = \int_{\lambda} I_{\text{-}}(\lambda) d\lambda /\int_{\lambda} I_0(\lambda) d\lambda$.  We first express the normalized spectra from Ref.~\cite{Kato2013} Fig. 4a as a linear combination of the normalized basis spectra, $\hat{I}_0(\lambda)$ and $\hat{I}_{\text{-}}(\lambda)$ (see Figs.~\ref{fig:NVminusTheoryVsExperiment} and~\ref{fig:NVzeroExperiment}): 
\begin{align}
\hat{f}'_{\text{NV}^{\text{-}}}(\lambda) &= \alpha \hat{I}_0(\lambda) + (1-\alpha) \hat{I}_{\text{-}}(\lambda) \label{fminusnorm} \\
\hat{f}'_{\text{NV}^0}(\lambda) &= (1-\beta) \hat{I}_0(\lambda) + \beta \hat{I}_{\text{-}}(\lambda). \label{fzeronorm}
\end{align}
We find $\alpha = 0.0853$ and $\beta = 0.273$. 
The non-normalized PL spectra can be written as
\begin{align}
f'_{\text{NV}^{\text{-}}}(\lambda) &= C \left[a_0 \hat{I}_0(\lambda) + a_{\text{-}} \kappa_{532}\hat{I}_{\text{-}}(\lambda) \right]\label{fminus} \\
f'_{\text{NV}^0}(\lambda) &= C \left[b_0 \hat{I}_0(\lambda) + b_{\text{-}} \kappa_{532}\hat{I}_{\text{-}}(\lambda) \right]\label{fzero} 
\end{align}
where $C$ is an overall scaling constant and the constants $a_0$, $a_{\text{-}}$, $b_0$, and $b_{\text{-}}$ satisfy
\begin{equation}
    a_0+a_{\text{-}} = b_0+b_{\text{-}}, \label{numberconserved}
\end{equation}
which should be interpreted as the requirement that the quantity [NV$^\text{-}$]+[NV$^0$] be independent of the applied voltage. From Eqns.~\ref{fminus}, \ref{fzero} we can rewrite Eqn.~\eqref{kappaprimeone} as 
\begin{equation}
    \kappa_{532}' = \frac{a_0 + a_{\text{-}} \kappa_{532}}{b_0 + b_{\text{-}} \kappa_{532}}. \label{kappaprimenew}
\end{equation}
Comparison of Eqns.~\ref{fminusnorm},~\ref{fzeronorm} with Eqns.~\ref{fminus},~\ref{fzero} allows us to note
\begin{align}
\alpha &= \frac{a_0}{a_0 +a_{\text{-}} \kappa_{532}}\label{alpha}  \\
\beta &= \frac{b_0}{b_0 +b_{\text{-}} \kappa_{532}}\label{beta}
\end{align}
Solving Eqns.~\ref{numberconserved},~\ref{kappaprimenew},~\ref{alpha},~\ref{beta} yields $\kappa_{532} = 2.17$. This value of $\kappa$ should be taken as a lower limit on the true value of $\kappa_{532}$ at low 532 nm optical excitation intensity since low optical excitation intensity was not employed in the work of Ref.~\cite{Kato2013}.

\begin{figure*}[thp]
    \centering
        \includegraphics[width=\textwidth]{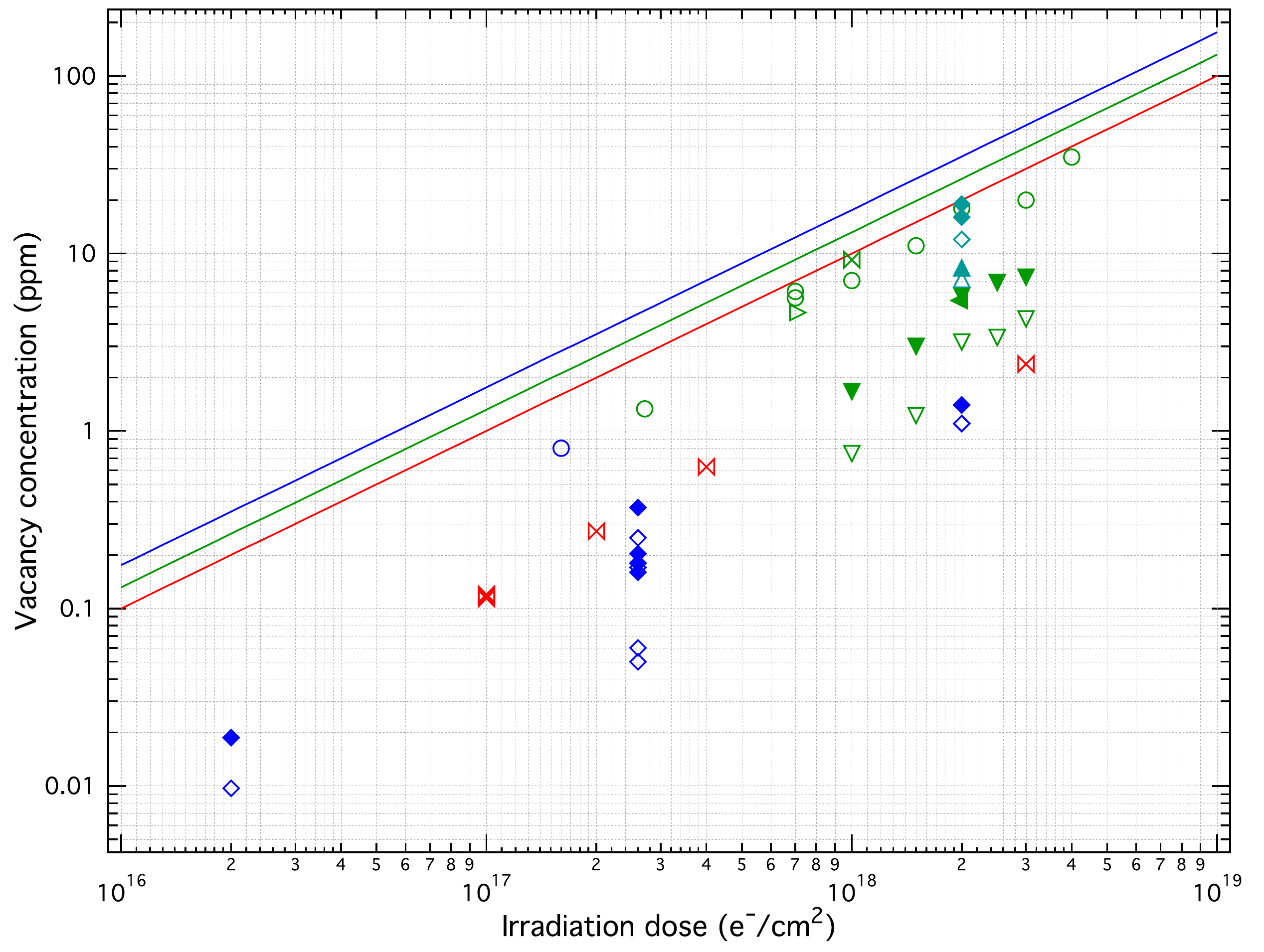}
        \caption{Reported values of monovacancy concentrations in ppm generated by electron irradiation at 1\,MeV ({\color{red}{\rule[.6mm]{3mm}{.3mm}}}), 1.9-2\,MeV ({\color{igorgreen}{\rule[.6mm]{3mm}{.3mm}}}), 3\,MeV ({\color{igoraquamarine}{\rule[.6mm]{3mm}{.3mm}}}), and 4.5-5\,MeV ({\color{blue}{\rule[.6mm]{3mm}{.3mm}}}) from the diamond literature, measured by 77\,K UV-Vis spectrophotometry using the oscillator strengths $f_\text{GR1}$ and $f_\text{ND1}$ reported in Ref.~\cite{twitchen1999correlation}. Solid lines represent the calculated dependence of vacancy creation on irradiation dose from SRIM calculations~\cite{ziegler2010SRIM} in Ref.~\cite{Campbell2000}, neglecting vacancy-interstitial recombination for electron irradiation at 1\,MeV (\textcolor{red}{$-$}), 2\,MeV (\textcolor{igorgreen}{$-$}), and 5\,MeV (\textcolor{blue}{$-$}). Filled markers denote measurements of the total monovacancy concentration $\text{V}^0 + \text{V}^\text{-}$ while open markers denote measurements of only V$^0$. All diamond samples included here contain nitrogen concentrations $\lesssim 10$\,ppm. The following markers denote measurements reported in the following references: \textcolor{red}{$\mathlarger{\mathlarger{\Join}}$} = this work at 1\,MeV, \textcolor{igorgreen}{$\mathlarger{\mathlarger{\mathlarger{\mathlarger{\mathlarger{\blacktriangleleft}}}}}$} = Ref.~\cite{lawson1998ontheexistence} at 1.9\,MeV recalculated using $f_\text{GR1}$ and $f_\text{ND1}$ from Ref.~\cite{twitchen1999correlation}, \textcolor{igorgreen}{$\mathlarger{\mathbin{\bigtriangledown}}$} and \textcolor{igorgreen}{$\mathlarger{\mathlarger{\mathlarger{\mathlarger{\mathlarger{\mathbin{\blacktriangledown}}}}}}$} = Ref.~\cite{twitchen1999correlation} at 1.9\,MeV,  \textcolor{igorgreen}{$\mathlarger{\mathlarger{\Join}}$} = Ref.~\cite{allers1998annealing} at 2\,MeV recalculated using $f_\text{GR1}$ and $f_\text{ND1}$ from Ref.~\cite{twitchen1999correlation},  \textcolor{igorgreen}{$\mathlarger{\mathlarger{\mathlarger{\mathlarger{\circ}}}}$} = Ref.~\cite{twitchen1999electronparamagneticresonanceandopticalabsorption} at 2\,MeV, \textcolor{igorgreen}{$\mathlarger{\mathlarger{\mathbin{\rhd}}}$} = Ref.~\cite{newton2002recombination} at 2\,MeV,  \textcolor{igoraquamarine}{$\mathlarger{\mathlarger{\mathlarger{\mathlarger{\mathlarger{\mathbin{\blacktriangleup}}}}}}$} and \textcolor{igoraquamarine}{$\mathlarger{\mathbin{\bigtriangleup}}$} = Ref.~\cite{collins2003production} at 3\,MeV, \textcolor{igoraquamarine}{$\mathlarger{\mathlarger{\mathlarger{\mathlarger{\mathbin{\blackdiamond}}}}}$} and \textcolor{igoraquamarine}{$\mathlarger{\mathlarger{\mathlarger{\mathlarger{\mathbin{\diamond}}}}}$} = Ref.~\cite{iakoubovskii2005evidence} at 3\,MeV, \textcolor{blue}{$\mathlarger{\mathlarger{\mathlarger{\mathlarger{\mathbin{\blackdiamond}}}}}$} and \textcolor{blue}{$\mathlarger{\mathlarger{\mathlarger{\mathlarger{\diamond}}}}$} = Ref.~\cite{E6patent2010WO149775} at 4.5\,MeV, and \textcolor{blue}{$\mathlarger{\mathlarger{\mathlarger{\mathlarger{\circ}}}}$} = Ref.~\cite{Fraczek2017} at 4.5\,MeV.     
        }
        \label{fig:Vacanciesvsdose}
\end{figure*}

\subsection{Annealing and nitrogen conversion study details}
\subsubsection{Irradiation studies in the literature}
\label{Appendix:IrradiationStudy}

This section reviews prior literature studies reporting monovacancy creation rates by electron irradiation.  Figure~\ref{fig:Vacanciesvsdose} displays vacancies measured in the diamond literature via UV-Vis absorption spectrophotometry for varying irradiation doses~\cite{lawson1998ontheexistence,twitchen1999correlation,allers1998annealing,twitchen1999electronparamagneticresonanceandopticalabsorption,newton2002recombination,collins2003production,iakoubovskii2005evidence,Fraczek2017,E6patent2010WO149775}, alongside simulated vacancy creation rates from Ref.~\cite{Campbell2000}. In type IIa diamonds with low concentrations of electron donors, irradiation primarily creates neutral monovacancies (V$^0$, GR1), whereas in diamonds with a moderate to high concentration of electron donors, such as type Ib diamonds, negative monovacancies (V$^\text{-}$, ND1) are created at similar or greater rates than neutral monovacancies~\cite{collins2003production}. Fig.~\ref{fig:Vacanciesvsdose} depicts only the V$^0$ concentration for the former sample type and the total vacancy concentration (V$^0$ + V$^\text{-}$) for the latter type. 

The GR1 center displays a zero-phonon absorption doublet at 740.9 nm and 744.4 nm~\cite{Zaitsev2001}, while the ND1 center displays a zero-phonon line at 393.6 nm~\cite{dyer1965irradiation}. For the data depicted in Fig.~\ref{fig:Vacanciesvsdose}, vacancy concentration is calculated by integrating under the absorption features to obtain $A_\text{GR1}$ and $A_\text{ND1}$, the absorption strengths in meV\,cm$^{-1}$, and then solving the equations $A_\text{GR1} = f_\text{GR1}[V^0]$ and $A_\text{ND1} = f_\text{ND1}$[V$^\text{-}$], where $f$ represents the transition’s oscillator strength~\cite{twitchen1999correlation,davies1999current}. The authors of Ref.~\cite{davies1992vacancy} measured the ratio $f_\text{ND1}/f_\text {GR1} = 4.0$ using detailed balance arguments applied to an annealing study. Later, the authors of Ref.~\cite{twitchen1999correlation} determined that $f_\text{ND1} = 4.8(2) \times 10^{-16}$ meV\,cm$^2$ by correlating ND1 absorption measurements with well-calibrated EPR measurements of [V$^\text{-}$]. Although V$^0$ is spinless and thus EPR inactive, $f_\text{GR1}$ is determined to be $1.2(3) \times 10^{-16}$ meV\,cm$^2$ by dividing $f_\text{ND1}$ by $f_\text{ND1}/f_\text {GR1} = 4.0$. All vacancy concentrations reported in Fig.~\ref{fig:Vacanciesvsdose} use these calibration constants. 

Large variation is apparent between data sets from different publications, and, more specifically, between different irradiation facilities. The diamonds from Refs.~\cite{twitchen1999electronparamagneticresonanceandopticalabsorption,newton2002recombination,allers1998annealing} were all irradiated at Reading University in the UK and all show significantly higher vacancy creation rates than diamonds irradiated at other facilities. Strikingly, Refs.~\cite{twitchen1999electronparamagneticresonanceandopticalabsorption} and~\cite{twitchen1999correlation}, published by the same group within the same year using approximately the same electron energy ($\approx\!2$~MeV), report vacancy creation rates of 0.50 cm$^{-1}$ and 1.53 cm$^{-1}$ respectively, a difference of $3\times$. This discrepancy leaves open the possibility of a calibration error in the applied dosage at one or more irradiation facilities, or a variable which is not controlled for. 

Furthermore, although Ref.~\cite{newton2002recombination} observed the vacancy creation rate to be independent of temperature up to $600$ K, such temperatures could be exceeded depending on the irradiation setup; inadequate heat sinking combined with high beam current could cause diamonds to be heated above the temperature at which vacancies become mobile and recombine with interstitials, thereby reducing the measured post-irradiation vacancy concentration for a given dose. The diamonds represented in Fig.~\ref{fig:Vacanciesvsdose} are irradiated with beam currents ranging ranging from $\upmu$A~\cite{twitchen1999correlation, twitchen1999electronparamagneticresonanceandopticalabsorption,allers1998annealing} to tens of mA~\cite{elementsixpersonalcommunication,prismgempersonalcommunication}.  In Ref.~\cite{E6patent2010WO149775}, the vacancies created per dose is seen to vary by more than $2\times$ between different diamonds. It is unclear whether this variation arises from differences in the samples~\cite{davies1992vacancy}, or inconsistency in the irradiation conditions (such as electron beam inhomogeneity). It is possible that more than one of these surmised explanations combine to describe the variation depicted in Fig.~\ref{fig:Vacanciesvsdose}. 
Moreover, as pointed out in Ref.~\cite{Collins2009}, there exists disagreement in the literature on the value of $f_\text{ND1}/f_\text {GR1}$. In particular, absorption measurements employing reversible charge interconversion between GR1 and ND1 find the ratio of oscillator strengths $f_\text{ND1}/f_\text{GR1}$ to range from 2 to 10~\cite{iakoubovskii2003annealing, deweerdt2007thesis,dyer1965irradiation}. This variation suggests a possible uncertainty in the y-axis of Fig.~\ref{fig:Vacanciesvsdose} of up to $\sim 5\times$. Such uncertainty complicates the determination of the required irradiation dose to create a fixed vacancy concentration, and thus generate a desired NV concentration after annealing.

We further note that all measured data points lie below the predicted relationship between dose and vacancy concentration from Ref.~\cite{Campbell2000}, which uses SRIM (Stopping and Range of Ions in Matter) calculations~\cite{ziegler2010SRIM}. These simulations ignore spontaneous recombination, which is crudely estimated to occur $30\%-50\%$ of the time for diamonds irradiated with $\sim\!2$ MeV electrons~\cite{Campbell2000,davies2001interstitials} and may occur at a higher rate for lower-energy electrons, where the vacancy-interstitial distance is reduced~\cite{davies2001interstitials}. Uncertainty in the value of $f_\text{ND1}/f_\text{GR1}$~\cite{collins2007optical,iakoubovskii2003annealing} could help account for the discrepancy between the simulations and much of reported data.

Indirect methods for determining the vacancy concentration produced by a given irradiation dose offer an independent alternative to UV-Vis absorption measurements. The indirect method presented in this paper finds a monovacancy creation rate of 0.52$\pm$0.26 cm$^{-1}$ for 1 MeV electron irradiation by Prism Gem. In comparison, direct vacancy concentration measurements by UV-Vis on five similar diamonds irradiated at the same facility (the $\mathlarger{\mathlarger{\Join}}$ symbols in Fig.~\ref{fig:Vacanciesvsdose}, excluding the point at $3\times 10^{18}$ e$^-/$cm$^{2}$), suggest a V$^0$ creation rate of $0.30\pm 0.05$ cm$^{-1}$ and no production of V$^\text{-}$ above the device's detection limit of [V$^\text{-}]\sim 100$ ppb. 
The indirect measurements detailed in the main text and these UV-Vis measurements are consistent to within the uncertainty in $f_\text{GR1}$, which demonstrates the need for additional detailed studies to reduce uncertainty and resolve the aforementioned discrepancies. 

The fit to Eqn.~\eqref{eqn:plversusdosage2} gives $B = 0.52\pm0.26$ cm$^{-1}$.

\subsubsection{Diffusion}\label{Appendix:Diffusion}

For one dimensional diffusion, the RMS distance $\sigma$ from the starting point after $N$ jumps of distance $L$ is given by $\sigma = L \sqrt{N}$. The diffusion coefficient is defined as $D \equiv \frac{\sigma^2}{2t_\text{ann}}$ where $t_\text{ann}$ is the anneal time during which the diffusion occurs. Combining the two equations allows the diffusion coefficient to be rewritten as $D = L^2\Gamma/2$ where $\Gamma\equiv N/t_\text{ann}$ is the vacancy jump frequency. It can be shown that the distance between two adjacent carbon atoms in the diamond lattice is $\frac{a\sqrt{3}}{4}$ where $a$ is the diamond lattice constant. Therefore, in one dimension $L = \frac{a}{4}$. Substituting in gives
\begin{equation}\label{DfromGammaa}
    D = \Gamma \frac{a^2}{32},
\end{equation}
which is consistent with alternative derivations~\cite{Pichler2004}. We note that the diffusion constant is independent of the dimensionality so the above diffusion constant holds for three dimensional diffusion in the diamond lattice~\cite{Pichler2004}. 

Comparison of samples with 1 hour post-irradiation anneal at 850~$^\circ$C with samples receiving a 1250~$^\circ$C post-irradiation anneal, suggests that 1250~$^\circ$C 1 hour anneal created approximately 4.5 times as many NVs as the 850~$^\circ$C anneal for irradiation doses $\sim \! 10^{16}$ e$^\text{-}$/cm$^2$ (see Fig.~\ref{fig:Fluorversusdosage}). Under the likely assumption that the 1 hour 1250~$^\circ$C anneal proceeded to completion (no monovacancies are left), this finding suggests that employing only the 1 hour 850~$^\circ$C anneal limits NV$^\text{T}$ to $\chi \approx 1/4.5$ of that expected from the introduced vacancy concentration alone. From the observed value of $\chi$ we can estimate the diffusion constant $D$ as we now describe. 

We assume vacancy diffusion can be modeled as a random walk on the diamond lattice and ignore sinks for vacancies other than substitutional nitrogen. We do not correct for any time the monovacancy spends in the negative charge state which is immobile at 850~$^\circ$ C~\cite{breuer1995ab,davies1992vacancy}. The number of unique lattice sites a vacancy is expected to visit in order to create an NV is 
\begin{equation}\label{eqn:need}
M_\text{need} =\frac{[\text{C}]}{4[\text{N}]F_c},
\end{equation}
where [C]=$1.76\times10^{17}$ cm$^{-3}$ is the density of carbon atoms in diamond and the factor $4$ accounts for the four lattice sites adjacent to a substitutional nitrogen, each of which will result in NV formation if visited by a vacancy. The number of unique sites visited is 
\begin{equation}\label{eqn:visited}
    M_\text{visited} = \Gamma t_\text{ann} \eta_\text{uni} = 32 D t_\text{ann} \eta_\text{uni}/a^2
\end{equation}
where $\eta_\text{uni}$ characterizes the ratio of unique sites visited to discrete steps during the random walk. Ref.~\cite{Fastenau1982} finds the fraction of unique sites visited for a random walk on a diamond lattice to be $n_\text{uni}\approx 0.56 n + 0.63\sqrt{n}$ where $n$ is the number of steps. This result is similar to the number of distinct sites visited after $n$ steps for other lattices where: s.c.~$n_\text{unique}\approx 0.66 + 0.57\sqrt{n}$, b.c.c.~$n_\text{unique}\approx 0.72 + 0.52\sqrt{n}$, and f.c.c.~$n_\text{unique}\approx 0.74 + 0.52\sqrt{n}$~\cite{Vineyard1963,Fastenau1982}. 

We assume the conversion efficiency of the annealing follows
\begin{equation}\label{eqn:annealingefficiency}
\chi = 1 - e^{-\frac{M_\text{visited}}{M_\text{need}}}.
\end{equation}
Combining Eqns.~\ref{eqn:need},~\ref{eqn:visited},~\ref{eqn:annealingefficiency} allows the diffusion constant to be written as
\begin{equation}
    D = -\frac{[\text{C}] a^2 \text{Log}[1-\chi]}{32 t_\text{ann} 4 [\text{N}] F_c \eta_\text{uni}}
\end{equation}
where $\chi \approx 1/4.5$ is the observed annealing efficiency for the 1 hour 850~$^\circ$C anneal. We obtain $D$ = 1.8 nm$^2$/s for monovacancies at 850~$^\circ$C, subject to the assumptions of our diffusion model. This value should be compared to the vacancy diffusion constant determined in Ref.~\cite{martin1999generation} which observes $D$ = 1.1 nm$^2$/s for annealing at 750~$^\circ$C and Ref.~\cite{acosta2009diamonds} which observes the upper limit of $D \leq 40$ nm$^2$/s for annealing at 1050~$^\circ$C.

\subsubsection{Comparison with crude diffusion theory}

A neutral monovacancy in diamond is surrounded by four carbon atoms. The average time for one of the carbon atoms to jump into the vacancy is given by~\cite{Fletcher1953}
\begin{equation}
\tau_\text{carbon} = \frac{e^{E_a/k_B T}}{\nu_0}
\end{equation}
where $E_a = 2.3 \pm 0.3$ eV~\cite{davies1992vacancy} is the activation energy, $k_B$ is the Boltzmann constant, $T$ is the temperature, and $\nu_0$ is the attempt frequency for an individual carbon atom. As there are four carbons, the average time for the vacancy to move one lattice site is 
\begin{equation}
\tau_\text{vacancy} = \frac{e^{E_a/k_B T}}{4\nu_0}
\end{equation}
so that we can write the vacancy jumping frequency as
\begin{equation}
\Gamma = 4\nu_0 e^{-\frac{E_a}{k_B T}}.
\end{equation}
Plugging the above into Eqn.~\eqref{DfromGammaa} gives
\begin{equation}
D=\frac{1}{8}\nu_0 a^2  e^{-\frac{E_a}{k_B T}}.
\end{equation}
Although the above equation is approximately correct, accurate values of the attempt frequency $\nu_0$ are difficult to determine a priori~\cite{swalin1973calculation}. In particular, Ref.~\cite{swalin1973calculation} notes, \textit{``For want of a suitable value, either the Debye frequency or Einstein frequency can be used but there is no fundamental reason why either is applicable''}. Nevertheless, employing the diamond Debye frequency of 40 THz for $\nu_0$ results in $D = 30$ nm$^2$/s, much higher than the experimental estimate of 1.8 nm$^2$/s. We note that the uncertainty on $E_a = 2.3\pm0.3$ results in a roughly 500$\times$ uncertainty in $D$ at 850~$^\circ$C, so the discrepancy between the experimentally estimated $D$ and that calculated from theory cannot be completely attributed to uncertainty in the value of $\nu_0$.

\subsection{$T_2$ and $T_1$ coherence study details}
\label{Appendix:T2T1}
\subsubsection{$T_2$ measurement protocol}
The $T_2$ coherence time for each diamond is measured as follows: \textcolor{black}{A moderate magnetic field of 40-60 gauss is applied along the ${<}$111${>}$ diamond crystallographic axis, allowing NV centers of only a single orientation to be addressed.} NV fluorescence collected from a $\pi/2-\pi-\pi/2$ spin-echo sequence and a subsequent $\pi/2-\pi-3\pi/2$ sequence is subtracted and normalized to eliminate common-mode laser intensity noise and other slow drifts.  To avoid anisotropic-hyperfine effects between the NV spin and the $^{13}$C nuclei~\cite{Stanwix2010}, special care is taken to minimize misalignment of the static magnetic field with the NV symmetry axis.  To achieve this condition, the bias magnetic field is aligned along one of the four diamond crystallographic axes, and only NVs parallel to this direction are addressed with resonant microwave (MW) radiation. The envelope of the spin-echo signal is modeled as a stretched exponential $e^{-(\tau/T_2)^p}$ with stretched exponential parameter $p$. While the collapses and revivals in the fluorescence signal due to the $^{13}$C nuclei are theoretically described by a fourth-power sinusoid~\cite{Laraoui2013} for a single NV spin, we instead used a linear combination of Lorentzian and Gaussian shapes for each revival to modulate the stretched exponential envelope.  The curve fits for our diamonds yielded $0.5 < p < 2$. From this fit we extract $T_2$.

\subsubsection{$T_1$ measurement protocol}
The longitudinal relaxation time $T_1$ is measured as follows: NVs are first optically initialized into the $m=0$ state.  Thereafter a near-resonant MW $\pi$ pulse transfers the population to the $m=1$ state, after which the population relaxes for a variable time $\tau$ toward the thermal equilibrium mixed state before being read out with a second optical pulse.  This yields the fluorescence trace $m_1$. The NV ensemble is optically initialized   and again allowed to relax again to thermal equilibrium with no MW pulse, giving fluorescence trace $m_0$. The population difference $m_0 - m_1$ is fit to an exponential decay with decay time $T_1$.  The subtraction of the signals eliminates common-mode noise from laser fluctuations during the measurement and the common fluorescence from the NV orientations not addressed by the MW $\pi$ pulse.

\end{document}